 \documentclass[preprint2]{aastex}

\usepackage{natbib}
\newcommand{\xmmnewton}{{\it XMM-Newton}}
\newcommand{\rosat}{{\it ROSAT}}
\newcommand{\bepposax}{{\it BeppoSAX}}
\newcommand{\einstein}{{\it Einstein}}
\newcommand{\chandra}{{\it Chandra}}
\newcommand{\bbxrt}{{\it BBXRT}}
\newcommand{\iras}{{\it IRAS}}
\newcommand{\ctb}{CTB~109}

\newcommand{\NH}{\mbox {$N_{\rm H}$}}
\newcommand{\nh}{\mbox {$N_{\rm H}$}}
\newcommand{\hi}{H\,{\sc i}}
\newcommand{\hii}{H\,{\sc ii}}
\newcommand{\sii}{S\,{\sc ii}}
\newcommand{\oiii}{O\,{\sc iii}}
\newcommand{\nex}{Ne\,{\sc x}}
\newcommand{\mgxi}{Mg\,{\sc xi}}
\newcommand{\mgxii}{Mg\,{\sc xii}}
\newcommand{\sixiii}{Si\,{\sc xiii}}
\newcommand{\sixiv}{Si\,{\sc xiv}}
\newcommand{\sxv}{S\,{\sc xv}}

\slugcomment{}

\shorttitle{\xmmnewton\ observations of CTB~109}
\shortauthors{Sasaki et al.}

\begin{document}


\title{\xmmnewton\ observations of the \\
    Galactic Supernova Remnant CTB~109 (G109.1--1.0)}


\author{Manami Sasaki} 
\email{msasaki@cfa.harvard.edu}

\author{Paul P. Plucinsky}

\author{Terrance J. Gaetz}

\author{Randall K. Smith}

\author{Richard J. Edgar}

\and

\author{Patrick O. Slane}

\affil{Harvard-Smithsonian Center for Astrophysics,
    60 Garden Street, Cambridge, MA 02138}


\begin{abstract}
We present the analysis of the X-ray Multi-Mirror Mission (\xmmnewton) 
European Photon Imaging Camera (EPIC) data of the Galactic supernova remnant 
(SNR) \ctb\ (G109.1--1.0). \ctb\ is associated with the anomalous X-ray 
pulsar (AXP) 
1E\,2259+586 and has an unusual semi-circular morphology in both the X-ray and 
the radio, and an extended X-ray bright interior region known as the `Lobe'. 
The deep EPIC mosaic image of the remnant shows no emission towards the west 
where a giant molecular cloud complex is located. No morphological 
connection between the Lobe and the AXP is found. We find remarkably little 
spectral variation across the remnant given the large intensity variations. 
All spectra of the shell and the Lobe are well fitted by a single-temperature 
non-equilibrium ionization model for a collisional plasma with solar 
abundances ($kT \approx 0.5 - 0.7$~keV, 
$\tau = \int n_{\rm e} dt \approx 1 - 4 \times 10^{11}$~s~cm$^{-3}$, 
$\NH\ \approx 5 - 7 \times 10^{21}$~cm$^{-2}$). 
There is no indication of nonthermal emission in the Lobe or the shell.
We conclude that the Lobe originated from an interaction of the SNR 
shock wave with an interstellar cloud. Applying the Sedov solution for the 
undisturbed eastern part of the SNR, and assuming full equilibration between 
the electrons and ions behind the shock front, the SNR shock velocity is 
derived as $v_{\rm s} = 720\pm60$~km~s$^{-1}$, the remnant age as 
$t = (8.8\pm0.9) \times 10^{3}~d_{3}$~yr, the initial energy as 
$E_{0} = (7.4\pm2.9) \times 10^{50}~d_{3}^{2.5}$~ergs, and the pre-shock
density of the nuclei in the ambient medium as
$n_{0} = (0.16\pm0.02)~d_{3}^{-0.5}$~cm$^{-3}$, at an assumed distance of
$D = 3.0~d_{3}$~kpc.
Assuming \ctb\ and 1E\,2259+586 are associated, these values constrain the age
and the environment of the progenitor of the SNR and the pulsar.
\end{abstract}




\section{Introduction}

The Galactic supernova remnant (SNR) \ctb\ (G109.2--1.0) 
was discovered in X-rays with \einstein\ by \citet{1980Natur.287..805G}. 
In the radio band, it was identified as an SNR
by \citet{1981ApJ...246L.127H} in the Galactic
plane survey at $\lambda$49~cm with the Westerbork Synthesis Radio Telescope.
\ctb\ is the host remnant of the anomalous X-ray pulsar (AXP)
1E\,2259+586 with a spin period of $P = 6.98$~s (Fahlman \& Gregory 
\citeyear{1981BAAS...13..533F}, \citeyear{1983IAUS..101..445F}) 
and a spin-down rate of 
$3 - 6 \times 10^{-13}$~s~s$^{-1}$ \citep{1992PASJ...44....9I}.
In 2002, an outburst of 1E\,2259+586 occurred with bursts similar to 
Soft Gamma Repeaters and with a sudden spin up of $\delta\nu/\nu = 4 \times
10^{-6}$
\citep{2003ApJ...588L..93K,2003astroph0310852G,2003astroph0310575W}.

The distance to the SNR has been estimated from various observations as 
3 -- 6~kpc \citep[][and references
therein]{1983PASJ...35..447S,1984ApJ...283..147H,2002ApJ...576..169K}.
A distance of $3.0\pm0.5$~kpc has been derived by \citet{2002ApJ...576..169K}
by measuring the spectroscopic distances and radial velocities of \hii\
regions and comparing those values to the radial velocity of \ctb\ measured by
\citet{1990ApJ...351..157T}.
The remnant is embedded in a large complex of \hii\ regions 
which extends over 400~pc along the Galactic plane.

\begin{figure*}
\centering
\caption{\label{color}
See 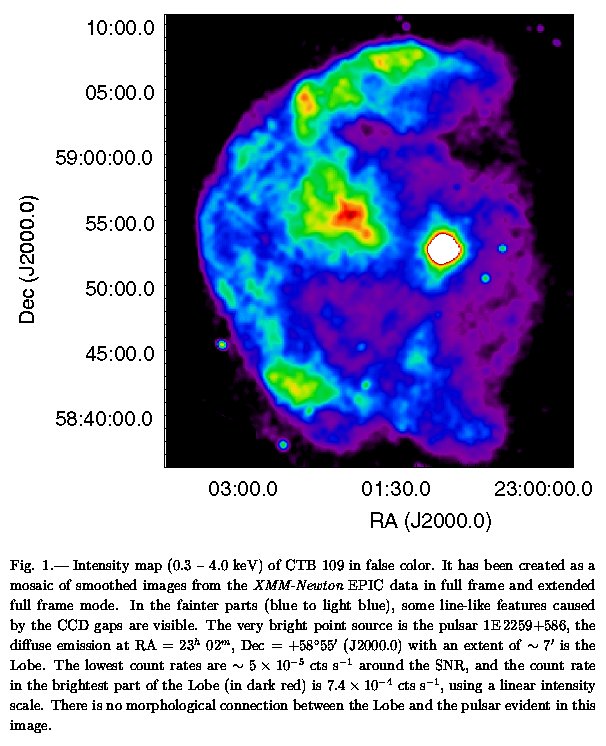.}
\end{figure*}

At radio frequencies, \citet{1984ApJ...283..147H} have derived a spectral 
index of $\alpha = 0.50\pm0.04$, 
for flux density $S_{\nu} \propto \nu^{-\alpha}$ and frequency $\nu$, 
which does not vary across the remnant. 
The remnant shell is incomplete in the west, both in radio and X-rays (see
Fig.\,\ref{color}).
Bright spots are found around the rim, but the 
features do not correlate in the two bands. 
No radio point source is found at the position of the X-ray pulsar, which is 
displaced by 3\farcm6 from the geometrical center of the radio shell. 

\ctb\ is located near a giant molecular cloud (GMC) complex
\citep{1980AJ.....85.1612I} which contains five \hii\ regions. 
\citet{1981Natur.293..549H} and \citet{1985PASJ...37..345T} have suggested 
that this GMC complex is associated with the SNR based on CO and $^{13}$CO
observations.
Data from the Position Sensitive Proportional Counter (PSPC) of the 
R\"ontgensatellit (\rosat) confirm that there is no X-ray emission in the 
west \citep[][hereafter RP97]{1997ApJ...484..828R}.
The SNR shell does not extend further to the west behind the GMC in either 
radio or X-rays; because absorption is negligible in the radio continuum,
this implies that the semi-circular shape of the remnant is not due to 
absorption. More likely, the SNR shock has been stopped by the 
GMC complex on one side. To the east, the interstellar medium (ISM) 
density appears to be lower, 
as indicated in the \hi\ map of \citet{2002ApJ...576..169K}.
In CO ($J = 1 - 0$) and $^{13}$CO ($J = 1 - 0$) observations,
\citet{1987A&A...184..279T} have found an arm-like CO 
ridge (`CO arm') which is anti-correlated with the X-ray emission.

\ctb\ has a strikingly bright region in X-rays in the interior, 
the `X-ray blob' or the `Lobe'.
\citet{1983IAUS..101..429G} have suggested that the AXP emits a jet and 
that the bright Lobe is material excited by a jet or the jet itself.
However, Rho \& Petre (\citeyear{1993AAS...18310107R}, 
\citeyear{1997ApJ...484..828R}) have shown that the \rosat\/ PSPC 
spectrum of the Lobe is thermal in origin and shows no
evidence of synchrotron emission, arguing against a jet interpretation.
\citet{1995MNRAS.277..549H} have found no morphological evidence to support a 
pulsar jet origin for the Lobe in the \rosat\/ High Resolution Imager (HRI) 
data and conclude that the bright X-ray emission from the Lobe is most 
likely caused by a density enhancement in material stripped from the 
cloud complex.
Finally, the \chandra\ image \citep{2001ApJ...563L..45P} reveals no 
morphological connection between the Lobe and the AXP. 
\citet{1997ApJ...484..828R} show that for parts of the shell, the 
X-ray spectrum is well fitted by a single-component thermal model, whereas 
for the Lobe and the northern and the southern part of the shell, two thermal 
components are required. 
A \bepposax\ Low-Energy Concentrator Spectrometer (LECS) spectrum of the whole 
SNR excluding the pulsar and the Lobe has been analyzed by 
\citet{1998A&A...330..175P}. The parameters of the best fit to the LECS data
with a non-equilibrium model are $kT = 0.95^{+0.65}_{-0.27}$~keV and the
ionization timescale
$n_{\rm e} t = 3.8^{+3.8}_{-1.6} \times 10^{11}$~cm$^{-3}$~s.

Surveys of Galactic SNRs have been performed with the Infrared Astronomical
Satellite (\iras), revealing infrared emission from \ctb\
\citep{1989ApJS...70..181A,1992ApJS...81..715S}. The data in the wavelength
bands of 12, 25, 60, and 100~$\mu$m show a dense dust cloud which is 
correlated with bright \hii\ regions, located in the south and west of the 
remnant \citep{1989MNRAS.238..649C}. 
Furthermore, an enhancement in infrared emission is found between the Lobe 
and the north-east shell. 

Optical emission is detected mainly near the bright northeastern and southern 
radio emission \citep{1981ApJ...246L.127H}. Spectra of filamentary structures 
in the south confirm their shock origin \citep{1981Natur.291..132B}.
H$\alpha$ and [\oiii] images of \citet{1995AJ....110..747F} reveal
previously unknown filaments, e.g. faint filaments near the projected center 
of the remnant, along the southwestern edge of the bright Lobe. From the
filament spectra, they derive a shock velocity of $\sim100$~km~s$^{-1}$ and
pre-shock cloud densities of 5 -- 20~cm$^{-3}$.

We have undertaken observations of \ctb\ with the \xmmnewton\
observatory to provide the deepest X-ray image of this SNR and to
provide the highest quality X-ray spectral data to date. 
The primary objectives of our analysis are to determine the 
evolutionary 
parameters of the remnant assuming it is in the Sedov phase
and to determine the nature of the Lobe emission. The deep
X-ray images are used to study the morphology of this remnant
in comparison to the data at other wavelengths and to search for
any morphological connection between the pulsar and the Lobe.
The high quality spectral data are used to determine the
evolutionary parameters of the remnant (age, shock velocity, ambient
density) and to search for spectral variations within the remnant.  
The evolutionary parameters determined from the X-ray spectral fits
provide useful constraints not only on the age and the
environment in which the remnant formed but also for the pulsar.
This is particularly useful for the AXP since there is no other reliable
technique for estimating the age of this object
\citep{2003ApJ...588L..93K,2003astroph0310575W}. The detailed spectral fits of
the shell and the Lobe provide useful insights into the nature of the
Lobe emission. In particular, the high-quality spectra of the Lobe
are used to determine if the abundances are enhanced (evidence of
ejecta), if there is a nonthermal component, and if the temperature
and ionization state of the plasma is similar to that of the shell.

\section{\xmmnewton\ Data}

We have observed \ctb\ with three pointings of the X-ray Multi-Mirror Mission 
\citep[\xmmnewton,][]{2001A&A...365L...1J,2000SPIE.4012..731A} in the AO1 
period covering the southern
(hereafter: pointing S), northern (pointing N), and eastern parts (pointing E). 
We analyze data from the European Photon Imaging Cameras (EPIC) 
PN \citep{2001A&A...365L..18S}, EPIC MOS1, and EPIC MOS2 
\citep{2001A&A...365L..27T}.
The exposure times, after removing flares and periods of high background, are 
$\sim$10~ks. 
Furthermore, we have access to observation 0038140101 (PI: V.\ Kaspi, 
hereafter pointing P1), with a net integration time of 34~ks for MOS1/2. 
We also analyze the target of opportunity (TOO) dataset of the pulsar 
1E\,2259+586 in the \xmmnewton\ Archive (Obs.\ ID 0155350301, hereafter P2) 
with 17~ks of MOS2 data
in full frame mode. The MOS1 was in partial window mode, covering only the 
pulsar. In observations P1 and P2, the PN was in small window mode. Therefore, 
we do not consider the PN data of the observation P1 and the PN and MOS1 data
of observation P2 in our analysis. Observation details of the datasets used
are listed in Table \ref{obstab}. 
Starting from the observational data files (ODFs), the data are processed with 
\xmmnewton\ Science Analysis System (XMMSAS) version 5.4.1.
For EPIC PN, only single and double pattern events are used, 
whereas for the MOS1 and 2, singles to quadruples are selected.

\subsection{Intensity Map}\label{intensmap}

The SNR has been fully covered in five pointings with different directions 
(three for the shell and two centered on the pulsar). We therefore merge
the images from the different pointings to obtain complete exposure corrected 
images of the remnant using the procedure {\tt emosaic}.
For all the pointings, events are selected in the energy bands: 
0.3 -- 0.9~keV, 0.9 -- 1.5~keV, 1.5 -- 4.0~keV, and 4.0~keV -- 12.0~keV.
Images are created with a bin size of 4\arcsec\ and smoothed with circular
Gaussian function that has a sigma of 3 pixels in both dimensions RA and Dec. 
In the 4.0~keV -- 12.0~keV images, only significant emission from the 
pulsar is seen, but not from the rest of the SNR. 

Figure \ref{color} shows a mosaic image in the energy band of 
0.3 -- 4.0~keV in false color. 
The remnant's X-ray emission fades off to the west, lacking a clearly defined 
edge. In the interior, the Lobe is brighter than any part of the shell and 
there are dark regions north and south of the Lobe. The Lobe emission is well
separated from the pulsar.
We estimate the distance of the shell from the pulsar in the mosaic image
by fitting a circle centered on the pulsar to each sector and obtain: 
18\farcm5 for the eastern shell, 17\farcm3 for the southern shell, and 
16\farcm6 for the northern shell. 
Thus, the eastern shell has the largest projected distance from the pulsar. 

\subsection{RGB Composite Image}\label{rgbima}

\begin{figure*}
\centering
\caption{\label{truecolor}
See 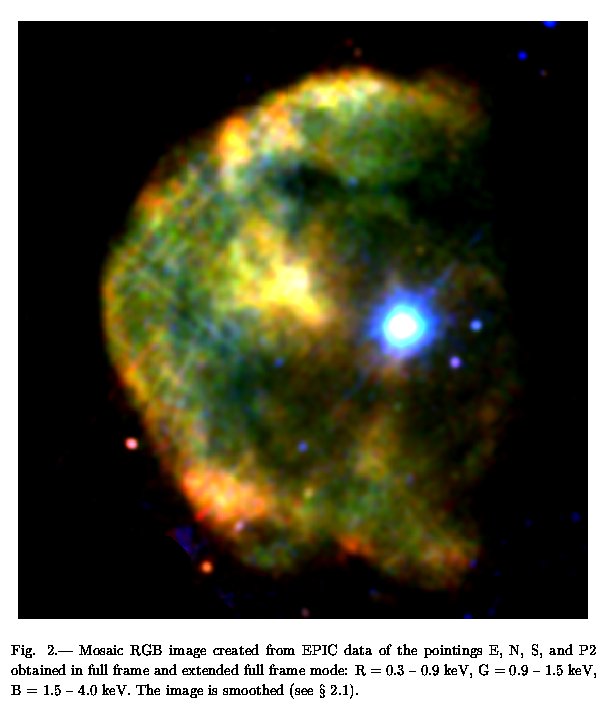.}
\end{figure*}

\begin{figure}
\centering
\includegraphics[width=\columnwidth]{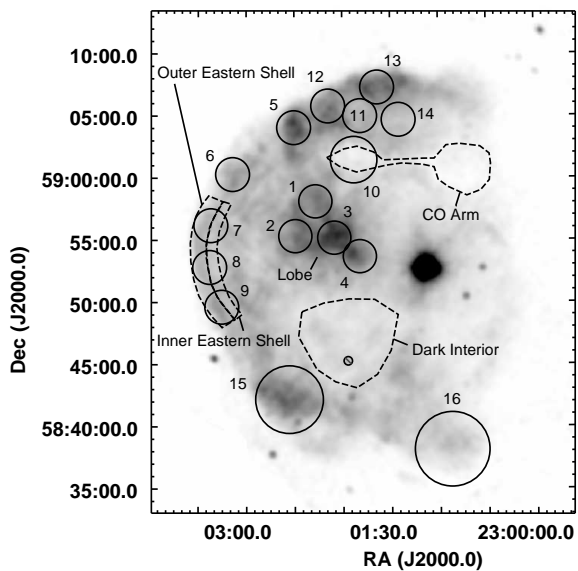}
\caption{\label{regions}
Mosaic of all five pointings towards \ctb\ 
in the energy band of 0.3 -- 4.0~keV with analyzed
regions. For circular regions, numbers as used in the paper are given, and 
the non-circular regions analyzed with {\tt evigweight} are shown with dashed
lines. A point source is excluded in the region `Dark Interior' (small 
canceled circle).}
\end{figure}

As no emission from the remnant is found above 4~keV except from the pulsar,
we create images in the energy bands below 4~keV in order to identify 
possible spectral variations within the remnant. The images in the 
bands red (R) = 0.3 -- 0.9~keV, green (G) = 0.9 -- 1.5~keV, and 
blue (B) = 1.5 -- 4.0~keV are combined into an RGB mosaic image 
(Fig.\,\ref{truecolor}). 
We then use this image to guide our selection of regions to be examined
more closely through detailed spectral fitting. The selected regions are shown 
in Figure \ref{regions}. 

The shell is slightly redder than the interior and has some red clumps. There
is a small sector of the shell with lower emission in the northeast, 
coinciding with a bright region in the radio continuum
\citep{2002ApJ...576..169K}. Next to this region, there are red knots in the
RGB composite image of the X-ray shell (regions 5 and 6 in Fig.\,\ref{regions}).
Furthermore, there is a larger red spot in the south (region 15). 
There are also color variations within the Lobe, the outskirts being redder
than the central part of the Lobe.

In addition to the very bright pulsar, fainter point sources are found in and 
around the remnant. Two point sources located southeast of the remnant appear
red in the RGB composite image. 
The one further east at RA = 23$^h$03$^m$19.6$^s$, 
Dec = +58\degr45\arcmin29\arcsec\ (J2000.0) is a star of the spectral type 
G5.5:V \citep[USNO-A2.0 1425-14515707,][]{1998yCat.1252....0M} and is
coincident with source No 3 in RP97. 
The other source at RA = 23$^h$02$^m$43.7$^s$, 
Dec = +58\degr37\arcmin46\arcsec\ (J2000.0) is coincident with source No 4 in
RP97; it is also coincident with a point source in the DSS image. Presumably, 
it is also a Galactic star. The point sources close to the pulsar (in 
projection) have blue or violet colors. The source at 
RA = 23$^h$00$^m$33.3$^s$, Dec = +58\degr52\arcmin45\arcsec\ (J2000.0)
located west of the pulsar is USNO-A2.0 1425-14436845 (No 1 in RP97), a
star of the spectral type G9.5:III-IV. The source at RA = 23$^h$00$^m$43.3$^s$, 
Dec = +58\degr50\arcmin28\arcsec\ (J2000.0) south of USNO-A2.0 1425-14436845,
is only known as an X-ray source (No 2 in RP97). There is an optical point 
source on the DSS image at the X-ray position.
Therefore, these hard point sources 
seem to be Galactic stars as well. Enhanced column density in the foreground 
or high intrinsic absorption seems to be responsible for the hard spectrum.
We extracted spectra for these point sources, but the low number of counts
does not allow detailed spectral fitting.

\subsection{Background}

Since \ctb\ is a relatively faint diffuse object, a good 
knowledge of the background is crucial for the spectral analysis. 
We studied the background properties of the EPIC cameras based on the
treatments of
\citet{2001XMM.CAL.TN0016L},
\citet{2002A&A...389...93L}, and
\citet{2003A&A...409..395R}.

In the case of a point source a local background extracted from another 
region of the same data as the source can be used. 
For extended objects, it is not possible to obtain background data at small 
distances from the source, i.e.\ a few arcseconds away from the source like for 
point or point-like sources. Instead one has to extract the background data 
far from the target, from a part of the detector that is not illuminated by 
source emission. In this case the distance between the source and the 
background region is large (few arcminutes) and the difference of the detector 
background at the two positions could become significant (see next paragraph). 
In the case of a diffuse source that fills 
the entire field of view, it is even impossible to estimate the local 
background from the same data. If one selects a background region far from the 
source itself, many effects can produce an inappropriate background due to the
different chip position:
First, the effective area of the mirrors depends on the off-axis angle.
Photons are subject to vignetting, but particles are not.
Second, high-energy particles which interact with material surrounding the
detector produce fluorescence, the magnitude of this fluorescence emission
can vary significantly with position on the detector, especially 
for the PN detector.
Third, the spectral response depends on the position on the detector, 
especially for the PN. 

The internal `quiescent' background of the EPIC cameras consists of 
fluorescence 
lines and events produced by charged particles. The fluorescence 
depends on the camera body materials and the incident flux of high energy
charged particles. Therefore, the fluorescent component varies with position
on the detector and with time. In addition, the high energy charged particles
can produce background events directly in the detector. Finally, low energy
protons can reflect off of the mirror surfaces and reach the detectors in the
focal plane. The background produced by low energy protons is highly variable
in time for the \xmmnewton\ instruments. 

There are two types of background template files available that can
be used for the analysis of EPIC data: `Closed' data and blank sky data.
`Closed' data have been obtained with the filter wheel in the `closed' 
position; these data include only the internal detector background. 
However, this 
instrument configuration is not representative of the usual observation 
where the CCDs are exposed to photons and low energy charged particles.
For the blank sky data, datasets for different pointings have been merged for 
each EPIC camera after eliminating sources from the data. These datasets
comprise the detector background and an average cosmic X-ray background.
Since the background component due to high energy charged particles is
included in these data at an unknown level, we need to model it using the 
spectral shape determined from the closed data.

\subsection{Extraction of EPIC Spectra}

\subsubsection{Circular Regions}\label{speccirc}

Although the cosmic X-ray background is a significant uncertainty in the blank 
sky data, we decide to use the blank sky instead of the `closed' data,
because the latter have poor photon statistics and are therefore not suitable 
for extracting small regions. 
We use the blank sky data of the background analysis group in Birmingham, UK
\footnote{See {\it http://www.sr.bham.ac.uk/xmm3/}.},
because they have data for both the medium and the thin filters, as well 
as for the full frame and extended full frame PN modes.
The Birmingham group have also developed scripts that can be used for
the background analysis.
The script {\tt skycast} converts the detector coordinates of the template 
background datasets into sky coordinates using the pointing direction of
the observation to be analyzed and adds sky coordinates to the 
background data.
The script {\tt createspectra} extracts spectra in specified circular (or 
annular) regions from the data and background-file. 
Ancillary response files (ARFs) and redistribution matrix files (RMFs) 
are produced for the corresponding detector regions using
the XMMSAS commands {\tt arfgen} and {\tt rmfgen}. Since the source spectrum 
and associated blank sky background spectrum are created at 
the same position of the detector,
the ARF and RMF are correct for both source and background. 

\begin{figure*}
\centering
\includegraphics[scale=0.41]{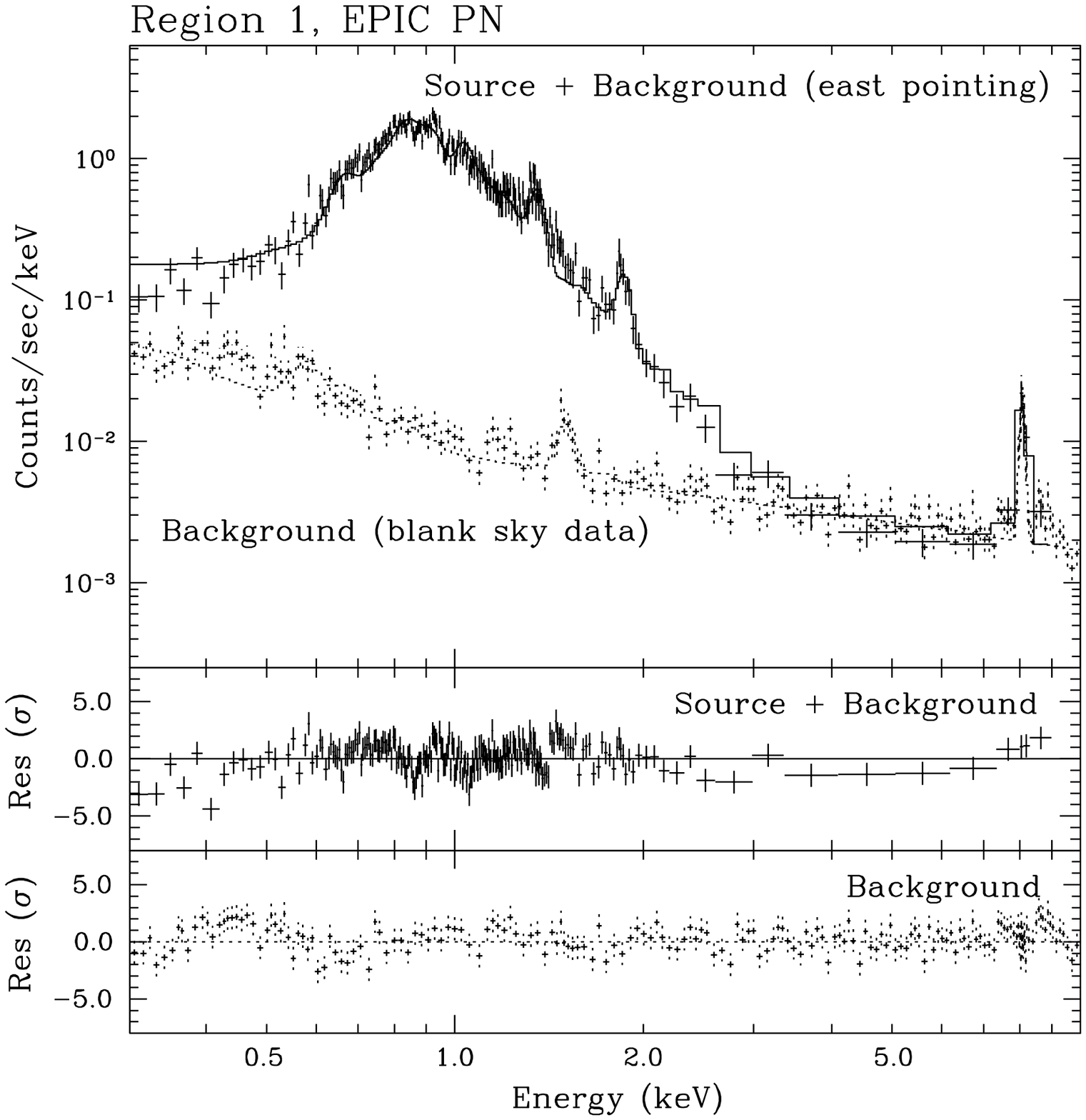}\hspace{4mm}
\includegraphics[scale=0.41]{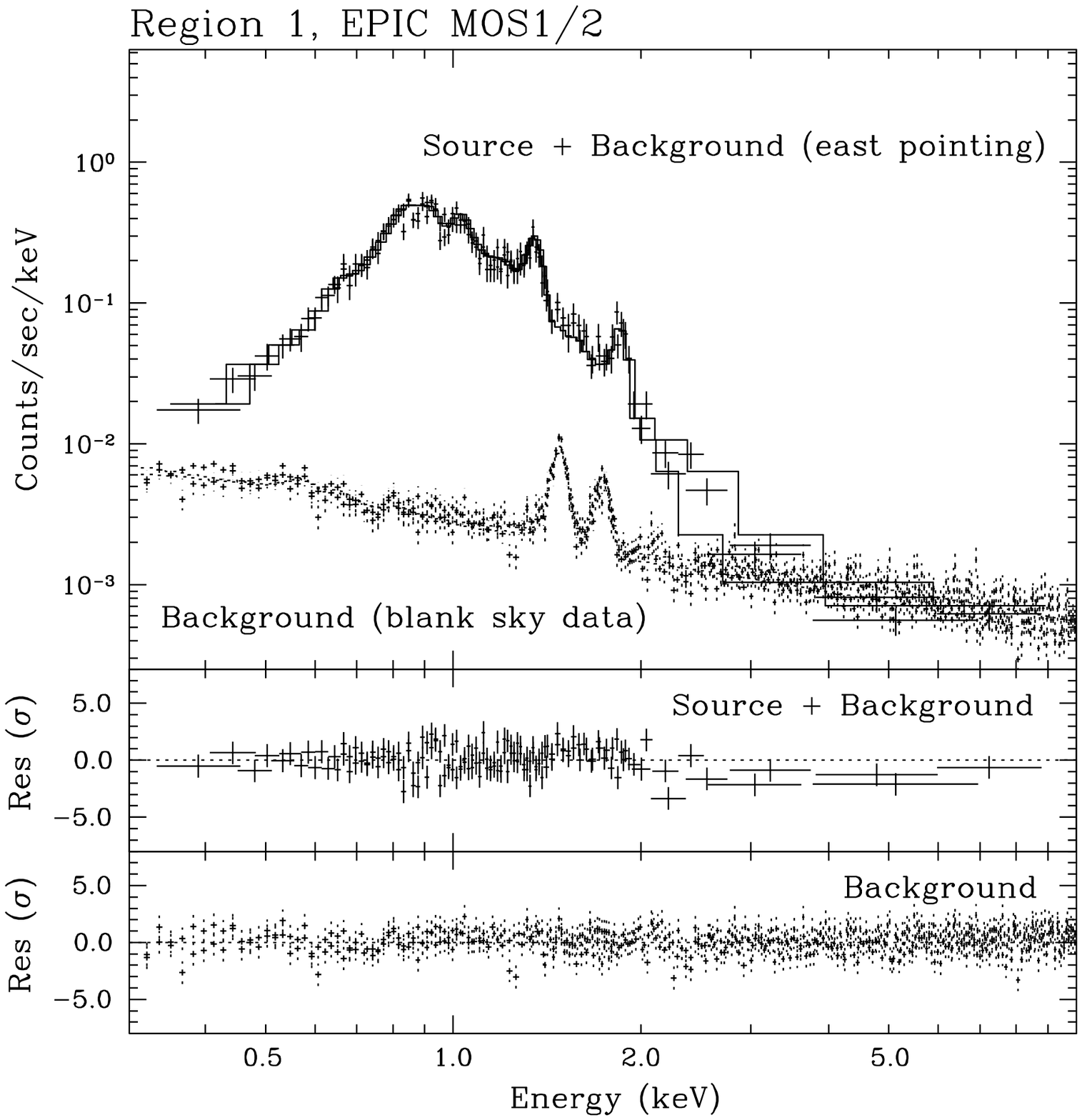}
\caption{\label{soubgspectra}
\xmmnewton\ EPIC spectra of the Lobe region 1 extracted from the pointing E 
data are shown with VNEI model fits for PN and MOS1/2. Solid line is used
for source + background spectrum, dashed line for background spectrum. 
The background spectrum has not been subtracted from the source spectrum, but 
has been modeled simultaneously and is included in the spectral model of the 
source spectrum. 
Strong fluorescence lines are visible in the background 
spectra: Al-K line at 1.5~keV and Cu-Ni-Zn-K line complex around 8~keV in the 
PN spectrum (left), and Al-K at 1.5~keV and Si-K at 1.7~keV in the MOS
spectra (right).
}
\end{figure*}

Spectra are analyzed with XSPEC Ver.\ 11.3. We read in 
the source and the blank sky background spectra as two 
datasets: source (1), background (2). We model the background spectrum as a 
combination of a powerlaw component and two Raymond \& Smith 
\citep[][hereafter RS]{1977ApJS...35..419R} components with temperatures 
$kT$ = 0.08~keV and 0.2~keV \citep{2001XMM.CAL.TN0016L,
2002A&A...389...93L}. Zero-width Gaussians are used to reproduce the 
fluorescence lines of the detector. The detector background is
modeled with a powerlaw component and Gaussians using the XSPEC option 
`/b' to suppress the application of the ARF.
An additional thermal component is used to model the SNR 
emission (set to zero for the background dataset).
In Figure \ref{soubgspectra}, the EPIC PN and MOS1/2 spectra of region 1
(see Fig.\,\ref{regions}) extracted from the E pointing are displayed
together with the corresponding blank sky background spectra. We show the 
spectrum from region 1 since this is the faintest part of the Lobe, and hence 
the most sensitive to the background subtraction. 

In order to convince ourselves that the background in the source spectrum can 
be well estimated from the blank sky data, we compare the spectra extracted as 
source (from observations) and background (from blank sky data) in a 
region outside the SNR. Both for PN and for MOS1/2, the off-remnant spectrum 
from observations matches the blank sky spectrum very well in the softer 
($< 0.6$~keV) and harder ($> 1.5$~keV) bands as well as for the fluorescence 
lines. There is a small excess in the off-remnant spectrum between $\sim0.6$ 
and 1.5~keV. This faint emission is well fitted with a thermal model 
component and is likely Galactic diffuse X-ray emission. Since the remnant 
spectrum dominates the background spectrum in the 0.6 -- 1.5~keV band 
(see Fig.\,\ref{soubgspectra}), using the blank sky data as background 
introduces only a small systematic uncertainty in the derived parameters for 
the remnant spectra. 
We verify this by modeling the spectra of regions in the 
remnant shell including an additional thermal component for the local diffuse 
X-ray background, as derived from the off-remnant spectrum. 
The parameters of the model component for the remnant emission 
(foreground absorption, temperature, and ionization timescale, 
see \S\ \ref{specana}) are consistent 
with results from the fit using only the blank sky data as background. The 
abundances of the elements with emission lines in this energy band (O, Ne, and 
Mg) also yield fit results that agree well with the results without the 
additional component, i.e.\ solar abundances. Consequently, the use of the 
blank sky data as background should have little effect on our determination of 
the neutral hydrogen column density and on the temperature of the thermal 
component.  

\subsubsection{Arbitrary Extraction Region Shape}

Since the existing version of {\tt arfgen} only computes ARF files
for circular and annular extraction regions, our analysis has been largely 
limited to circular regions. 
However, the XMMSAS command {\tt evigweight} calculates the vignetting
correction for each event and thus one can use the on-axis ARF. 
In order to subtract the non-vignetted events from the
observation data, it is better to use the `closed' data for 
the background, as this dataset only consists of the non-vignetted component,
which is the internal detector background, and thus includes no 
additional external background emission. Compared to the blank sky data,
the `closed' data have far poorer statistics. Therefore, we use this method 
only for non-circular extraction regions.

After computing the weight factor for the vignetting correction for each event,
spectra are extracted from the new event files and RMFs are created with
{\tt rmfgen}. The source data still include the non-vignetted events which
have not passed through the X-ray telescope, but now have a weighting factor 
applied. This misapplication of the weighting factor to non-vignetted events 
can be compensated for by applying the same weights to the `closed' background
data. Coordinates of the `closed' data
are converted into the corresponding sky position using {\tt skycast}. The 
vignetting correction weights are computed for the `closed' data.
Finally, spectra are extracted in the same regions as for the observed source 
data. Again the background spectrum is used as second dataset and modeled as in 
\S\ \ref{speccirc}.

\subsubsection{Comparison of the Two Methods}

In order to check for consistency between the two methods for creating spectral
data, spectra of the Lobe regions (see \S\ \ref{sectlobe}), 
extracted from the pointing E data, are analyzed using both methods and 
compared. For the first method, we use the blank sky data as background data 
and create the ARF with {\tt arfgen}. For the second method, the event files
are vignetting-corrected with {\tt evigweight} and on-axis ARF and `closed'
background data are used. 
All the fitted parameters derived from the two methods agree within 90\% 
confidence limits. Therefore, we are confident that
either method may be used without introducing systematic errors which 
dominate over the statistical errors. 

\section{The Spectral Analysis}\label{specana}

The \xmmnewton\ EPIC data allow us to perform spatially resolved spectral 
analysis of the SNR. Extraction regions are shown in Figure \ref{regions} on 
the 0.3 -- 4.0~keV mosaic image. Datasets and extraction regions are selected 
as follows:
\begin{itemize}
\item 
As we have shown in \S\ \ref{intensmap}, the eastern part of the shell 
is farthest from the pulsar. Since it is located opposite to the GMC complex in 
the west, the eastern shell segment most likely indicates a blast wave  
propagating into a lower-density ISM \citep{2002ApJ...576..169K}.
For the spectral analysis, we cover the eastern shell with three circular
regions (7 -- 9) and also arc regions for comparison.
In addition, we extract spectra for regions next to the radio bright
knot in the northeast (regions 5 and 6). Dataset E is used for all these 
analyses.
\item 
The shell is brightest in the northern part; it has been covered completely
by pointing N. We select circular regions from this dataset centered on
the brighter knots (regions 12 and 13) and also analyze interior regions
close to the northern shell for comparison (regions 11 and 14). 
\item 
In the southern part of the shell which does not have as clearly a defined
edge as the eastern or the northern portions of the shell, we study the EPIC 
spectra (using observation S) for two regions (15, 16) with soft emission 
(see RGB composite image, Fig.\,\ref{truecolor}).
\item 
In CO observations, an enhancement of CO gas has been found which extends
from the GMC in the west across the northern part of the remnant
\citep{1987A&A...184..279T}. This dense
material is called the CO arm. Its position and shape coincides well with an 
X-ray faint interior region, north of the pulsar and the Lobe 
(see Fig.\,\ref{color}). 
Spectra are analyzed for a large circular region (10) as 
well as for a non-circular region which follows the shape of the very low  
emission region. Data from observations N and P1 are used. 
\item
The interior of the remnant has remarkably low surface brightness in the 
south (`Dark Interior'). This region is analyzed using the P1 data which have 
the longest exposure time.
\item 
We extract spectra for four regions of the Lobe 
(regions 1 -- 4 in Fig.\,\ref{regions}) from the observations E, P1, 
and P2, in order to look for spectral variations.
\end{itemize}
All the EPIC spectra are analyzed in XSPEC using an energy range of 
0.3 -- 10.0~keV. The spectra are grouped with a minimum of 20 counts per bin
and the $\chi^{2}$ statistic is used.
The errors quoted are 90\% confidence limits.
As can be seen in Figure \ref{soubgspectra}, the emission above $\sim4$~keV
is mainly background emission and is crucial for estimating the background.
Although we select the energy range of 0.3 -- 10.0~keV for spectral fitting, 
we will only show the spectrum below 4~keV in the following figures in order 
to clearly display the SNR emission. 

For analyzing the \xmmnewton\ EPIC spectra we use the following models: 
VMEKAL, which describes an emission spectrum for hot
ionized gas in collisional ionization equilibrium with variable abundances 
\citep{1985A&AS...62..197M,1986A&AS...65..511M,1992SRON..........K,1995ApJ...438L.115L},
and VNEI, which is a model for a non-equilibrium ionization (NEI) collisional 
plasma, again with variable abundances
\citep{1983ApJS...51..115H,1994ApJ...429..710B,1995ApJ...438L.115L,2001ApJ...548..820B}.
The VNEI model includes a parameter for the ionization timescale, 
$\tau = \int n_{\rm e} dt$, an integral over the time since the gas was 
shocked. 
In addition, we modeled some regions in the shell with the model 
VPSHOCK for a constant temperature, plane-parallel shock which includes a range 
of ionization timescales, parameterized by the lower limit and the upper limit
of the $\tau$ range
\citep[][and references therein]{2001ApJ...548..820B}.

\begin{figure}
\centering
\includegraphics[width=\columnwidth]{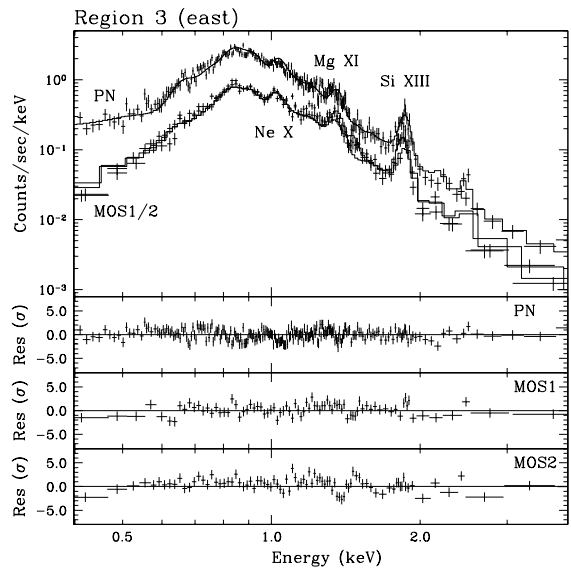}
\caption{\label{lobespectra}
\xmmnewton\ EPIC spectra of the Lobe region 3 extracted from the pointing E 
data, with VNEI model fits. The position of the \nex, \mgxi, and \sixiii\ lines 
is indicated. Instead of subtracting the background spectrum from the source 
spectrum, it has been 
modeled simultaneously. The displayed source spectrum and its fitted model 
include the background spectrum components.
Background spectra are not plotted in this and the following spectrum figures.
}
\end{figure}

The \xmmnewton\ EPIC spectra of region 3 in the Lobe extracted from the 
pointing E data (with VNEI model fits) are presented in 
Figure \ref{lobespectra}. Emission line features corresponding to the 
\mgxi\ triplet and the \sixiii\ triplet are easily visible. Moreover, the 
\nex\ Ly$\alpha$ line can be clearly seen in the MOS data, 
as the energy resolution of MOS is slightly better than that of PN.

We first fit the spectra with free abundances. For all the spectra,
the abundances for all the elements are consistent with solar abundances.
We thereupon fix the abundances to solar values in order to reduce the number
of free parameters and hopefully to better constrain the remaining free
parameters. Only Mg and Si are still kept free, because these abundances have 
appeared to differ slightly from solar values for some regions and the line 
features are evident in the EPIC spectra. We allowed the Si and Mg abundances
to vary independently. For other elements like Ne or S, it is too difficult to 
distinguish between line and continuum emission in the corresponding energy 
intervals.

\begin{figure}
\centering
\includegraphics[width=\columnwidth]{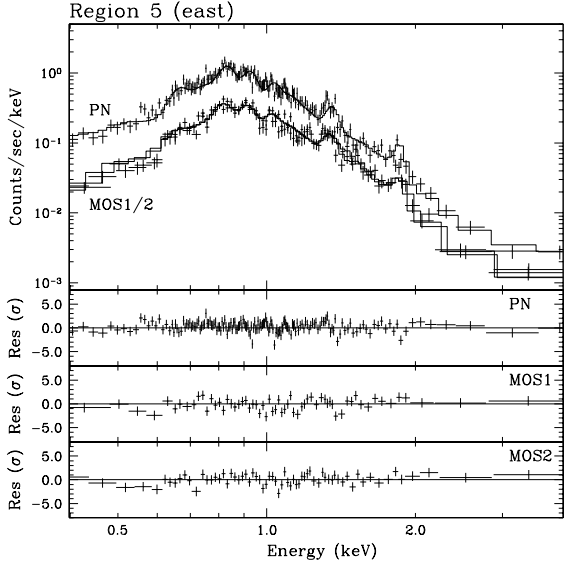}
\caption{\label{eastshellspectra}
\xmmnewton\ EPIC spectra of the region 5 in the eastern part of the shell 
extracted from the pointing E data, with VNEI model fits. 
For background spectrum see the caption of Figure \ref{lobespectra}.}
\end{figure}

\begin{figure}
\centering
\includegraphics[width=\columnwidth]{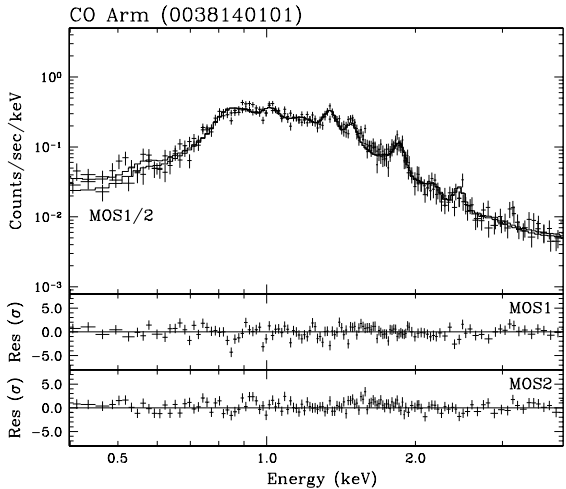}
\caption{\label{coarmspectra}
\xmmnewton\ EPIC MOS1/2 spectra of the CO arm (from pointing P1) with VNEI 
model fit. 
For background spectrum see the caption of Figure \ref{lobespectra}.}
\end{figure}

All the EPIC spectra are fitted better by a single-temperature
VNEI model than by a collisional ionization equilibrium VMEKAL model. 
In the following, we present only the VNEI results for all the regions, 
and discuss the VPSHOCK results for shell spectra where we expect the least 
disturbed plane-parallel shock. Tables \ref{mostab} and \ref{evigwtab} 
summarize the results of the spectral analysis of MOS1/2 data, using the VNEI 
model. Spectra of some selected regions with the best fit VNEI model are 
displayed in the following figures: Figure \ref{lobespectra} shows the 
PN and MOS1/2 spectra of the brightest part of the Lobe (region 3), 
the spectra in Figure \ref{eastshellspectra} are extracted from a bright
region in the shell (region 5), and Figure \ref{coarmspectra} shows the
heavily absorbed spectra of the CO arm. 
In all these plots, the non-background-subtracted source spectra are shown,
but not the simultaneously modeled background spectra.

The parameters of the VNEI model for the Lobe data of the pointing E, with the 
spectra from the two EPIC detector types PN and MOS fitted separately, 
are consistent with the MOS results of the Lobe regions in Table \ref{mostab} 
except that the Mg and Si abundances from the PN fits turn out to be sub-solar 
in some regions. 
The most likely explanation for this discrepancy between PN and MOS is 
the calibration issue addressed in the EPIC calibration status documentation
of April 2003 \citep{2003XMM.CAL.TN0018K}:
In the range from 0.3 -- 1.0~keV, the PN flux is up to 10\% higher than the
MOS flux, whereas above 1.5~keV, the MOS flux is up to 10\% higher. 
In the EPIC spectra, most of the emission lines cannot be resolved due to 
moderate spectral resolution. This complicates the determination of the
continuum. Assuming the same temperature and absorption, the flux inconsistency 
between the MOS and the PN can mimic lower abundances for the PN data for
elements like Mg and Si with lines above $\sim1$~keV, because the unresolved
lines cannot be distinguished from the continuum. Although most of the
fit parameters (absorption, temperature, ionization timescale) for the 
PN, MOS1, and MOS2 spectra are consistent, the discrepancy between the
abundances makes us conclude that the results from the MOS data (MOS1 and MOS2 
data fitted simultaneously, if data of both instruments are available) are 
more reliable than the PN results in terms of abundances. 
Therefore, we list the MOS1/2 results in Tables \ref{mostab} and \ref{evigwtab}.

\subsection{Shell}

\subsubsection{Eastern Part of the Shell}

The eastern part of the remnant shell is farthest from the pulsar 
1E\,2259+586 and has a more circular shape. 
The density of the ambient ISM is 
apparently lower on this side \citep{2002ApJ...576..169K}.
Thus the eastern portion of the shell seems to be 
the best region to study the least perturbed shock-ISM interaction.

The spectral analysis of regions in the eastern shell (regions 6 - 9) 
yields a temperature of $\sim0.6$~keV for the VNEI model,
which is slightly higher than those in the Lobe 
and the northern portion of the shell, though not significantly. The Mg and Si
abundances are consistent with solar, and the
ionization timescale is $\sim1 \times 10^{11}$~s~cm$^{-3}$,
similar to that in the northern part of the shell 
(see \S\ \ref{northshell}). The absorbing column density is \NH\ = 
$5 - 7 \times 10^{21}$~cm$^{-2}$. 
We also analyze the spectra of the 
regions in the eastern shell using the model VPSHOCK. The values for \nh\,,
$kT$, and the abundances agree well with the results of the VNEI model
in all four regions. The lower limit of $\tau$ has been set equal to zero.
The upper limit of the ionization timescale $\tau$ is
$3.4^{+0.7}_{-0.3} \times 10^{11}$~s~cm$^{-3}$ for region 6,
$4.8^{+7.2}_{-2.5} \times 10^{11}$~s~cm$^{-3}$ for region 7,
$3.5^{+30.0}_{-1.7} \times 10^{11}$~s~cm$^{-3}$ for region 8, and
$6.8^{+10.0}_{-1.1} \times 10^{11}$~s~cm$^{-3}$ for region 9, all higher than
and therefore consistent with the average values obtained from the VNEI model.

No indication of nonthermal emission is found in these spectra. 
We model the spectra with an additional powerlaw component and obtain an 
upper limit for the flux of the powerlaw component. Studies of 
nonthermal emission from the shells of SNRs yield photon indices of 
around 2.0: $\Gamma$ = 1.8 for Cas A 
\citep[below 16~keV,][]{1997ApJ...487L..97A} and $\sim2.2$ for SN 1006 
\citep{2003ApJ...589..827B,2004AdSpR..33..440A}. 
Therefore, we assume a photon index of $\Gamma = 2.0$ and obtain upper
limits of the nonthermal flux (0.3 -- 10~keV, 90\% confidence level):
$1.7 \times 10^{-14}$~ergs~cm$^{-2}$~s$^{-1}$ in region 6,
$1.3 \times 10^{-13}$~ergs~cm$^{-2}$~s$^{-1}$ in region 7,
$3.0 \times 10^{-14}$~ergs~cm$^{-2}$~s$^{-1}$ in region 8, and
$1.5 \times 10^{-14}$~ergs~cm$^{-2}$~s$^{-1}$ in region 9,
i.e. $<$0.2\%, $<$1.6\%, $<$0.4\%, and $<$0.2\% of the unabsorbed VNEI 
flux, respectively. \citet{2001ApJ...552L..39G} have studied the emission from
the forward shock in the northern part of Cas A and report that the flux of
the nonthermal emission is 50 -- 70\% of the total 0.5 -- 10.0~keV flux. 
In comparison, the upper limit of the nonthermal emission from the eastern 
part of the shell of \ctb\ in the energy range of 0.5 -- 10.0~keV is 
0.2 -- 1.4\%, about two orders of magnitude lower than in Cas A.

In radio continuum, \ctb\ has a very bright extended region approximately 
where the connecting line from the pulsar through the Lobe intersects the 
shell in the northeast \citep[see e.g.\ Fig.\,1 in][]{2002ApJ...576..169K}.
In region 5 which lies north of this radio bright region, the column density
\NH\ ($\sim8 \times 10^{21}$~cm$^{-2}$) and the ionization timescale $\tau$ 
($6 \times 10^{11}$~s~cm$^{-3}$ from the joint fit of PN and MOS1/2 data,
$4 \times 10^{11}$~s~cm$^{-3}$ from the fit of MOS1/2 data, see 
Table \ref{mostab}) are higher compared to the eastern shell regions south of 
the radio bright region, whereas the temperature is significantly lower: 
$kT = 0.29\pm0.03$~keV at the 90\% level, as obtained from both the PN and
MOS1/2 fit and the MOS1/2 fit. The $kT$ value for region 5 differs
from that in other regions also at the 99\% confidence level. 
The spectra are shown in Figure \ref{eastshellspectra}.
Fitting a model similar to other shell spectra (\nh\ = $6 \times
10^{21}$~cm$^{-2}$, $kT = 0.6$~keV, $\tau = 1 \times 10^{11}$~s~cm$^{-3}$)
results in a modeled spectrum with lower flux below 0.9~keV and flatter 
continuum above 2~keV, compared to the observed spectrum. 
The optical image of \citet{1995AJ....110..747F} 
shows that this region covers two bright optical filaments with a
[\sii]/H$\alpha$ ratio typical for shocked gas rather than of photoionized gas; 
they derive a pre-shock density of 10 -- 
20~cm$^{-3}$, higher than for other optical filaments seen in \ctb\,. 
This might indicate that a shock-cloud interaction is occurring in region 5. 
The ionization timescale for this region is $\sim$ 4 -- 6 times higher than 
in other parts of the remnant shell. Since the regions are all located at
similar radii, the time that has elapsed after these regions have 
interacted with the shock front can be assumed to be comparable. Therefore, 
the higher $\tau$ in region 5 is consistent with the shock having encountered 
a density enhancement.

Although we assume that the undisturbed eastern part of the remnant can be 
reproduced by a model based on the Sedov-Taylor-von Neumann similarity solution 
\citep{1959book..........S,1950PRSLA..201..159T,1947LASLTS.......vN},
the fit results are not satisfactory if we use the SEDOV model 
\citep{2001ApJ...548..820B} in XSPEC. This is presumably due to 
inhomogeneities in the ISM that result in flux and spectral variations in the 
remnant. Therefore, we focus on smaller shell segments
and analyze arc-shaped regions of the eastern shell (see
Fig.\,\ref{regions} and Table \ref{evigwtab}) using the {\tt evigweight} 
method. As the VNEI fit results show, the spectra in the `outer'
and `inner' shell seem to be different: The best fit temperature is 
slightly higher in the
outer shell, whereas the ionization timescale is lower. The foreground
absorption is lower for the outer shell than for the inner shell. 
However, if we compare the confidence contours of \nh\ vs.\ $kT$ and
$kT$ vs.\ $\tau$ for the two shell regions, the 90\% confidence regions
are fully separated, but the 99\% regions overlap. Thus, there is only a
marginal inconsistency between the temperatures and the ionization timescales 
of the outer shell and the inner shell.

For a Sedov-phase remnant expanding into a homogeneous medium,
the temperature and ionization timescale both increase radially inward.
This is an idealized case, however. If the $kT$ and $\tau$ differences 
are real, we may be seeing the effects of nonuniformities in the 
ambient medium.  If the shock (as seen in projection) encountered a 
density enhancement, that could cause a larger ionization timescale and 
lower temperature, and reduce the flux in the inner shell region relative
to the outer shell region.

\subsubsection{Northern Part of the Shell}\label{northshell}

For the northern portion of the shell (regions 12, 13)  
\NH\ $ \approx 7 \times 10^{21}$~cm$^{-2}$, $kT \approx 0.5$~keV, 
and $\tau \approx 1 \times 10^{11}$~s~cm$^{-3}$.
The temperature seems to be lower and 
the column density \nh\ higher than in the eastern portion of the shell.
For regions inside the northern shell segment (regions 11, 14), the
column density \nh\ seems to be slightly higher 
($8 - 9 \times 10^{21}$~cm$^{-2}$), perhaps indicating a gradual
increase of the foreground absorption in the direction of the CO arm.

\subsubsection{Southern Part of the Shell}

In the southern portion of the shell, 
there are two diffuse spots that appear red in the 
RGB composite image (Fig.\,\ref{truecolor}): a bright region in the southeast
(region 15) and the western tip of the shell in the south (region 16). 
The temperature is 0.6~keV and the ionization timescale 
$\tau \approx 1 \times 10^{11}$~s~cm$^{-3}$, comparable to the eastern part
of the shell. Rather than being a result of a lower temperature plasma, the 
red color of these regions appears to be a result from a lower \nh\ 
($\sim5 \times 10^{21}$~cm$^{-2}$, as derived from a joint fit of PN and 
MOS1/2 data, $\sim5.5 \times 10^{21}$~cm$^{-2}$, from MOS1/2 data only).

\subsection{CO Arm}

We analyze a circular region covering the extended tip of the 
CO arm (region 10). The column density \NH\ is higher than in the eastern 
part of the shell and the Lobe,
\nh\ $> 8 \times 10^{21}$~cm$^{-2}$, but the temperature, the Mg and Si 
abundances, and the ionization timescale are comparable to the shell of
the remnant ($kT = 0.5 - 0.6$~keV, $\tau = 1 - 2 \times
10^{11}$~s~cm$^{-3}$, as derived from PN and MOS1/2 data).

We also study a non-circular 
X-ray faint region corresponding to the CO arm (see Fig.\,\ref{regions}),
using {\tt evigweight}. Figure \ref{coarmspectra}  
shows the spectra of this region obtained with the EPIC MOS1/2 of the 
pointing P1; the flux falls off more rapidly below 1~keV compared to other 
regions of the SNR, consistent with the presence of additional absorption.
The result of the spectral fit of the MOS1/2 data is given in Table 
\ref{evigwtab};
it confirms that the column density \NH\ is almost $10^{22}$~cm$^{-2}$, 
higher than in other regions of the remnant. The intrinsic spectrum seems to 
be the same as in most of the other regions of the remnant.
Therefore, we conclude that this region appears fainter because of additional 
absorption.

\subsection{Dark Interior}

There is a low surface brightness region south of the Lobe inside the remnant 
shell. No enhanced CO emission is found in the observations by 
\citet{1990ApJ...351..157T} that coincide with this X-ray faint region. We 
extract the X-ray spectrum of this region using the {\tt evigweight} method
excluding the point source which is located within this region.
The parameters $kT$, Mg and Si abundances, and $\tau$ of the dark region 
are all consistent with the parameters of the CO arm (Table \ref{evigwtab}), 
indicating that the instrinsic spectra of these two regions are similar. Only 
the column density \nh\ is higher in the CO arm. The spectral parameters
also agree well with the parameters of region 1 (northern tip of the Lobe) 
which has about the same distance to the pulsar as this region. Although \nh\ 
does not differ significantly from that in region 1, it seems to be higher
in this dark region, indicating a higher absorption.

\subsection{Lobe}\label{sectlobe}

There are three observations which we can use to study the Lobe: pointings
E, P1, and P2. 
The results from pointings E and P2 are consistent with that from pointing P1.
The largest number of counts is obtained from the observation 
P1, because the exposure time is 2 to 3 times longer than in other 
observations. We also fit all the spectra (PN and MOS1/2 data of pointing E,
MOS1/2 data of pointing P1, and MOS2 data of pointing P2) simultaneously for
each of the regions 1 -- 4. The results of these simultaneous fits are all 
consistent with the results of the fit of the MOS 1/2 data of pointing P1 
(shown in Table \ref{mostab}). 

In all the Lobe regions, the best fit temperatures, $kT$, range from 
0.53 to 0.59~keV (from simultaneous fit of PN and MOS1/2 as well as from
the fit of MOS1/2 data, see Table \ref{mostab}).
For Lobe regions 2 -- 4, the column density \NH\ $\approx 5 \times
10^{21}$~cm$^{-2}$ and the ionization timescale 
$\tau = 2 - 3 \times 10^{11}$~s~cm$^{-3}$.
The rather small differences in temperature, ionization timescale, and \nh\ 
seem to be responsible for the apparent spectral differences seen as color   
variations in the RGB composite image (Fig.\,\ref{truecolor}).
But the detailed spectral analysis indicates that there are no statistically
significant differences among the four regions.
In the bright central part of the Lobe, Si appears to be slightly 
overabundant ($1.6\pm0.2$ times solar), while the Mg abundance is 
consistent with solar. This is not what we would expect if
the by-products of dust destruction were contributing to the X-ray
emitting material, because the Mg and Si abundances should track each other
\citep[][and references therein]{1989PASJ...41..853I,1994ApJ...431..188V}.
Since the other three regions have Si abundances consistent with solar and
all regions have Mg abundances consistent with solar, there is no compelling 
evidence for ejecta material in these regions. Furthermore, the global NEI 
fits that have all other elemental abundances set to solar values produce 
good fits with no large residuals around elemental emission lines.

\subsubsection{Line Emission Analysis}

All the EPIC spectra for the Lobe show line features which can be attributed to
\mgxi, \mgxii, and \sixiii\ lines, whereas the \sixiv\ line is too faint to be 
visible. Therefore, we use the \mgxi\ triplet and \mgxii\ Ly$\alpha$ line as
line diagnostics of the NEI conditions of the plasma.
We analyze the Lobe spectra from MOS1/2 data of pointing P1 and from MOS2 data 
of pointing P2. The spectrum is modeled
using a modified `APEC' \citep{2001ApJ...556L..91S} model that excludes all
emission lines\footnote{Contact R.\ Smith for a copy of this model.} for the 
continuum and Gaussians for the lines, over the energy range of 
1.15 -- 3.0~keV. The fitted parameters for the Gaussians are used to determine
the line centroids and fluxes.
The line energies for the Gaussians are modeled as follows:
1.211~keV for \nex\ Ly$\beta$, $\sim$1.340~keV for \mgxi\ triplet, 
1.472~keV for \mgxii\ Ly$\alpha$, $\sim$1.850~keV for \sixiii\ triplet, 
and $\sim$2.440~keV for \sxv\,.

With the spectral resolution of the EPIC detectors, we are not able to resolve 
the lines in the \mgxi\ triplet. Therefore, we model the \mgxi\ line 
feature with one Gaussian and determine the line centroid. 
Variations in the line centroid from region to region would be indicative 
of different relative strengths of the resonance, intercombination, and 
forbidden lines. However, the line centroids of \mgxi\ are equal for all four 
Lobe regions within the  
90\% confidence intervals (Table \ref{mglineratio}). The data simply are not of 
high enough quality to determine such a small shift. However, the EPIC 
spectral resolution is sufficient to resolve the \mgxi\ triplet from the 
\mgxii\ Ly$\alpha$ line. From the fluxes of the Mg line complexes, the ratio 
\mgxi\ triplet/\mgxii\ Ly$\alpha$ is calculated. In Table \ref{mglineratio}, 
the line energy in keV and the flux in photons/cm$^{2}$/s/extraction area are 
given for the \mgxi\ triplet and the \mgxii\ Ly$\alpha$ line, 
as well as the derived ratio of \mgxi\ triplet/\mgxii\ Ly$\alpha$.

\begin{figure}
\centering
\includegraphics[width=\columnwidth]{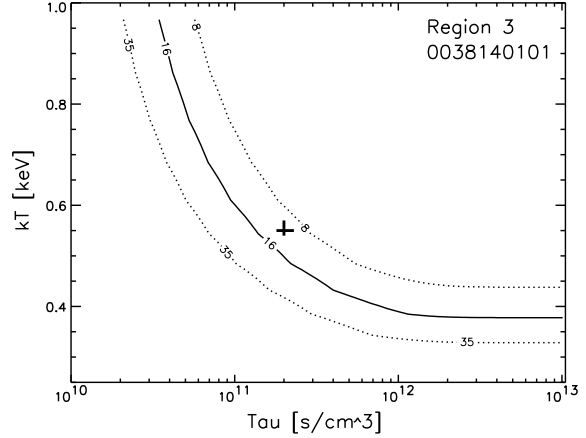}
\caption{\label{mgtriplyacont}
The line ratio \mgxi\,/\mgxii\ for region 3 derived from the spectral fits 
of P1 data is 
plotted as function of $\tau$ and $kT$ as solid line. The estimated range for
the error is shown with dotted lines. $kT$ and $\tau$ values resulting from 
the global spectral analysis (0.3 -- 10.0~keV with VNEI model) are plotted
with 90\% confidence errors. }
\end{figure}

The resulting lines fluxes and line ratios are compared to NEI plasma models
obtained using {\tt neiline} (see Appendix \ref{neilineapp}).
The calculation provides line emissivities for different temperatures and 
ionization timescales. 
In Figure \ref{mgtriplyacont}, the \mgxi\ triplet/\mgxii\ Ly$\alpha$ line 
ratio of region 3 for P1 data derived from the spectral analysis and error 
estimate based on the 90\% confidence intervals for the line fluxes is
plotted as a function of the ionization timescale $\tau$ and 
temperature $kT$. We also plot the best fit values from the global VNEI fit 
for $kT$ and $\tau$ and the 90\% errors. Since region 3 has the best 
statistics, it gives the narrowest band in the $kT$-$\tau$ diagram.
Within the errors, the VNEI value lies within the allowed region 
resulting from the line ratio analysis. From the diagram, 
we can derive the $kT$ range for region 3 as: $kT > 0.32$~keV. 
This lower limit is consistent with the 
results from the VNEI fit of the EPIC spectra in the total energy range of 
0.3 -- 10.0~keV (Table \ref{mostab}). 
The ratio of the \mgxi\ triplet to \mgxii\ Ly$\alpha$ does not provide a 
constraint on the ionization timescale.

These results show that line diagnostics are possible with CCD spectra. 
Compared to global fits, they suffer from fewer systematic uncertainties, 
while it is difficult to quantify the systematic errors of global fits. Line 
diagnostics are independent of foreground absorption and offer an additional 
method to constrain the plasma parameters. In our analysis the spectral 
resolution and the photon statistics only allow to set a lower limit on the 
temperature. However, the results obtained from fitting the lines only and 
from fitting the entire EPIC spectrum are consistent with each other. This 
method is a promising technique of analyzing plasma conditions and can be 
applied for brighter SNRs as well as for future missions with higher spectral 
resolution and larger collecting area.

\subsubsection{Nonthermal Emission}

In order to check if there is nonthermal emission from the Lobe 
that would be associated with a pulsar jet, 
we fit the spectra of regions 1 -- 4 in the energy range of 0.3 -- 10.0~keV 
with a model including a VNEI component and
an additional powerlaw component for the Lobe emission. The fit does not 
improve significantly, and the flux of the powerlaw component is consistent
with zero. Therefore, we derive an upper limit for the flux of the
powerlaw component, assuming a photon index of $\Gamma = 1.5$; typical
photon indices for pulsar jets are $\sim$1.2 -- 1.7 
\citep{2002ApJ...568L..49L,2002ApJ...569..878G,2003ApJ...591.1157P}. 
The nonthermal flux of the Lobe (0.3 -- 10~keV) is
$< 9.2 \times 10^{-15}$~ergs~cm$^{-2}$~s$^{-1}$ in region 1 
(90\% confidence level),
$< 8.0 \times 10^{-15}$~ergs~cm$^{-2}$~s$^{-1}$ in region 2,
$< 7.0 \times 10^{-15}$~ergs~cm$^{-2}$~s$^{-1}$ in region 3, and
$< 6.1 \times 10^{-15}$~ergs~cm$^{-2}$~s$^{-1}$ in region 4,
i.e. $<$0.05\%, $<$0.08\%, $<$0.05\%, and $<$0.07\% of the unabsorbed VNEI 
flux in regions 1 -- 4, respectively. For the entire Lobe, the upper limit
for the nonthermal emission is $6.9 \times 10^{-14}$~ergs~cm$^{-2}$~s$^{-1}$, 
corresponding to 
$8.2 \times 10^{-6}$~photons~keV$^{-1}$~cm$^{-2}$~s$^{-1}$ at 1~keV. 
For comparison, the intensity of the Vela pulsar jet
is $\sim4 \times 10^{-5}$~photons~keV$^{-1}$~cm$^{-2}$~s$^{-1}$ at 
1~keV \citep{2003ApJ...591.1157P}. 

\section{Estimates Based on the Sedov Solution}\label{sedovcalc}

The morphology of \ctb\ and the existence of the giant molecular cloud complex 
in the west suggests that the shock wave expanded into a
dense cloud and decelerated very quickly. To the east, 
the shock wave is expanding into a lower-density medium. 
\ctb\ seems to be confined by denser material in the north and the south
\citep{1995AJ....110..747F}, which might be responsible for the smaller 
radii of the northern and the southern X-ray shells compared to that of 
the eastern shell (\S\ \ref{intensmap}). 
However, the difference is relatively small ($\sim$5 -- 10\%) and overall 
the X-ray shell of \ctb\ can be considered as a semi-circle.
Calculations for the evolution of SNRs adjacent to a molecular cloud show 
that the shock propagation on the low-density side is almost unaffected
by the presence of the molecular cloud 
\citep[e.g.][]{1985A&A...145...70T,1992ApJ...388..127W} and can be modeled
as a spherical problem.
The X-ray temperature of $\sim3 - 8 \times 10^{6}$~K and the bright and 
well-defined X-ray rim of the remnant, as well as the lack of optical emission 
except in a few particular regions with higher density, indicate that the 
remnant has not yet reached the radiative phase.
To estimate physical parameters for the remnant, we apply the shock jump
conditions and the Sedov-Taylor-von Neumann similarity solution 
\citep{1959book..........S,1950PRSLA..201..159T,1947LASLTS.......vN}.
We assume a distance, $D$, to \ctb\ of $3.0\pm0.5$~kpc as estimated by 
\citet{2002ApJ...576..169K}, and introduce a scaling factor 
$d_{3} = D / 3.0$~kpc.
For the mass and densities, the following relations apply:
Assuming cosmic abundances \citep{1989GeCoA..53..197A}, 
the gas density (excluding electrons) is $n = 1.1~n_{\rm H}$, where 
$n_{\rm H}$ is the atomic H number density. The electron number density 
depends on the ionization state of the gas; if we assume at least single
ionization for each element, $n_{\rm e} = (1.1 - 1.2)~n_{\rm H}$, and the
total number density, including electrons, is 
$n_{\rm tot} = n_{\rm e} + n = (2.2 - 2.3)~n_{\rm H}$. The 
corresponding mass per free particle is $\bar{m} = (0.64 - 0.61)~m_{\rm p}$, 
with proton mass $m_{\rm p} = 1.67 \times 10^{-24}$~g. The mean mass per 
nucleus in any ionization stage
is $\bar{m}_{\rm n} = 1.4~m_{\rm p}$.
In the following, suffix `0' denotes pre-shock values, and suffix `s' 
denotes post-shock values. 

For the radius of the blast wave, we use the distance of the eastern 
shell from the pulsar as estimated from the EPIC image:
$R_{\rm s} = 18\farcm5\pm1\farcm0$ = $(16\pm1)~d_{3}$~pc
= $(5.0\pm0.3) \times 10^{19}~d_{3}$~cm.
If a supernova occurs near a molecular cloud, the shock can be reflected
at the cloud surface and cause the center of divergence of 
the velocity field to move away from the cloud \citep{1985A&A...145...70T}.
The pulsar might also have a proper motion. Consequently, in the case of \ctb\ 
the distance of the eastern part of the shell to the pulsar is only an 
estimate of the radius. 
An alternative way to estimate the radius of the remnant is to measure the
half of the extent of the shell in the north-south direction. However, as
there is dense matter both north and south of the remnant that might interact
with the shock front, the north-south extent rather provides a lower limit for 
the size of the remnant. 

For the temperature, we use the result from the VNEI fits of the spectra in 
regions 7 to 9. The weighted mean of the temperature are calculated
from the results which have been obtained by fitting the PN and MOS1/2 spectra 
separately, and from the results of the simultaneous fits of the PN and MOS1/2 
spectra. As the comparison of these results with the values for the eastern
shell (Table \ref{evigwtab}) show, there is a discrepancy of 
$\la$0.08~keV between the average temperature derived from the 
circular regions and the temperature obtained for the outer shell. 
To allow for the variation between the $kT$ estimates, we increase the error 
estimate to 0.10~keV for this calculation, and 
obtain for the temperature of the eastern part of the remnant shell:
$T_{\rm X} = 0.62\pm0.10$~keV = $(7.2\pm1.2) \times 10^{6}$~K.

At the discontinuity of the shock front, the density jump is $n = 4~n_{0}$, 
with $n_{0}$ being the pre-shock ambient density of nuclei. 
Inside the shock front, the density decreases towards
the center because of the adiabatic expansion of the remnant
\citep{1964ApJ...140..470H,1982ApJ...253..268C}.
In order to derive further parameters, we use the distribution of the
normalized density of the nuclei, $n(R)/n_{0} = n_{\rm H}(R)/n_{\rm H,0}$,
as calculated by \citet{1982ApJ...253..268C} 
and compute the emissivity numerically.
The emitting volume is modeled as a cylinder intersecting a spherical remnant 
with the long axis of the cylinder running along the line of sight through the 
remnant. The radius of the cylinder is simply the radius of the extraction 
region which, in this case, is $r = 80\arcsec = 1.2~d_{3}$~pc for
$D = 3.0~d_{3}$~kpc. 

From the spectral fits of the EPIC data of regions 7 to 9 with the VNEI model,
we derive the normalization in XSPEC as $K = \frac{10^{-14}}{4 \pi D^{2}}  
\times \int 1.2~n_{\rm H,s}^{2}~dV = (1.6\pm0.5) \times 10^{-3}$~cm$^{-5}$.
Since the integral of the normalized density $n(R)/n_{0}$ over the 
volume can be calculated numerically, the ambient ISM density $n_{0}$ is
estimated from the normalization $K$:
\begin{eqnarray}\label{xspecnorm}
K & = & \frac{1.2 \times 10^{-14} n_{\rm H,0}^{2}}{4 \pi D^{2}} 
\int \biggl( \frac{n_{\rm H}}{n_{\rm H,0}} \biggr)^{2} dV.
\end{eqnarray}
The projected volume is $V = 42~d_{3}^{3}$~pc$^{3}$,
and the integral is $\int (n_{\rm H}/n_{\rm H,0})^{2}~dV = 
240~d_{3}^{3}$~pc$^{3}$.
Thus the pre-shock H density is 
$n_{\rm H,0} = (0.14\pm0.02)~d_{3}^{-0.5}$~cm$^{-3}$, and 
the pre-shock density of nuclei
$n_{0} = 1.1~n_{\rm H,0} = (0.16\pm0.02)~d_{3}^{-0.5}$~cm$^{-3}$.

In the case of full equilibration between the electrons and the nuclei,
the shock velocity is related to the post-shock temperature as:
\begin{equation}\label{tempvel}
T_{\rm X} \approx T_{\rm s} = \frac{3 \bar{m}}{16 k} v_{\rm s}^{2},
\end{equation}
where $k = 1.38 \times 10^{-16}$~ergs~K$^{-1}$ is the Boltzmann's constant.
With a mean mass per free particle of $\bar{m} = 0.61~m_{\rm p}$ for a fully
ionized plasma, the shock velocity is estimated as $v_{\rm s} = 
[(16 k T_{\rm X})/(3 \times 0.61 m_{\rm p})]
^{\frac{1}{2}} = 720\pm60$~km~s$^{-1}$.

It is possible that the electrons and ions are not fully equilibrated in the
shock, in which case, the ion temperature $T_{\rm ion}$ may be higher, and the 
electron temperature $T_{\rm e}$ much lower than the mean plasma temperature
$T \approx T_{\rm ion}$. 
The electrons and ions would then equilibrate slowly through Coulomb 
collisions. The plasma X-ray emissivity depends primarily on the electron 
temperature, and the deduced velocity would be larger. 
\citet{2003ApJ...590..833G} and \citet{2003ApJ...590..846R} have analyzed the 
blast wave of the SNR DEM L71 in the Large Magellanic Cloud and find values for 
$T_{\rm e}/T_{\rm ion}$ at different locations ranging from 0.01 (almost
no equilibration) to 1.0 (full equilibration). 
The evolution of the electron temperature and its ratio to the mean post-shock 
temperature in two-fluid phase SNR shocks has been studied by 
\citet{1978PASJ...30..489I}. Applying the Sedov similarity solution to the
energy equation and the equation of state for the electron gas and assuming
Coulomb interactions between the electron and the ion gas, he shows
that the ratio $g = T_{\rm e}/T$ is a function of a reduced time 
variable $\nu = t_{3}(n_{0}^{8}/E_{51}^{3})^{1/14}$ which describes the 
thermal structure of the blast wave; $t_{3}$ is the time elapsed since the 
explosion (in units of $10^3$~yr), $n_{0}$ the pre-shock density of the
nuclei (in units of cm$^{-3}$), and $E_{51}$ is the initial blast energy (in 
units of $10^{51}$~ergs).
From X-ray measurements one obtains the X-ray temperature, $T_{\rm X}$,
the ambient density, $n_{0}$, and the radius, $R_{\rm s}$ of the SNR. Since 
these three values are related to each other as $f = T_{\rm X}/T = 
0.043 T_{\rm X} (R_{\rm s} n_{0})^{-1/2} \nu^{7/5}$, 
\citet{1978PASJ...30..489I} derives the value of $f = T_{\rm X}/T$ from 
the intersection of the curve $f(\nu)$ with the theoretical curve for Coulomb
equilibration $g(\nu) = T_{\rm e}/T$. 
Based on this method, we determine a lower limit for the ratio 
$T_{\rm e}/T_{\rm s} \approx T_{\rm X}/T_{\rm s} \approx 0.4$, and
an upper limit for the velocity $v_{\rm s} = [(16 k T_{\rm X})
/(3 \times 0.4 \times 0.61 m_{\rm p})]^{\frac{1}{2}} 
= 1140\pm90$~km~s$^{-1}$.

The age of the remnant can be estimated from the shock velocity using the
similarity solution:
\begin{equation}\label{simvel}
v_{\rm s} = \dot{R_{\rm s}} = \frac{2 R_{\rm s}}{5 t}.
\end{equation}
Under the assumption of full equilibration of electron and ion temperatures, 
this yields an age estimate of 
$t = (2.8\pm0.3) \times 10^{11}~d_{3}$~s
= $(8.8\pm0.9) \times 10^{3}~d_{3}$~yr.
If equilibration in the shock is incomplete, the age estimate would be lower.
For the lower limit of $T_{\rm e}/T_{\rm s} = 0.4$, the age would be
$t = (1.8\pm0.2) \times 10^{11}~d_{3}$~s
= $(5.6\pm0.6) \times 10^{3}~d_{3}$~yr.

We estimate the initial energy of the explosion from
\begin{equation}\label{radiussim}
R_{\rm s} = \Biggl(\frac{2.02 E_{0} t^{2}}{\bar{m}_{\rm n} n_{0}} \Biggr)^{\frac{1}{5}},
\end{equation}
where the mean mass of the nuclei is $\bar{m}_{\rm n} = 1.4~m_{\rm p}$.
Solving for $E_{0}$ gives
$E_{0} = \frac{1}{2.02} R_{\rm s}^{5} \bar{m}_{\rm n} n_{0} t^{-2} 
= (7.4\pm2.9) \times 10^{50}~d_{3}^{2.5}$~ergs, assuming full
equilibration between the electrons and ions,
and $E_{0} = (18.5\pm7.3) \times 10^{50}~d_{3}^{2.5}$~ergs, 
assuming partial equilibration with a lower limit of the temperature ratio
of $T_{\rm e}/T_{\rm s} = 0.4$.
The mass swept up by the SNR shock wave is 
$M = \frac{4 \pi}{3} R_{\rm s}^{3} 1.4 m_{\rm p} n_{0}
= (97\pm23)~d_{3}^{2.5}~M_{\sun}$ if we assume a uniform ambient medium.

\citet{1992ApJ...388..127W} have numerically modeled an SNR next to a dense 
molecular cloud, using the \citet{1960SPD.....5...46K} approximation. 
They obtain a geometry comparable to that of \ctb\,, 
$1.3 \times 10^{4}$~yr after the explosion in an interstellar
medium with a density of $n_{0} = 0.13$~cm$^{-3}$. This value for the 
ambient density agrees well with the result we obtain from the 
\xmmnewton\ data. The assumed initial energy is $E = 3.6 \times 10^{50}$~ergs, 
2 to 5 times lower than the energies we derive in our calculations.
Consequently, the age estimate of \citet{1992ApJ...388..127W} is higher.

\citet{1997ApJ...484..828R} have fitted a two-temperature Raymond \&
Smith model to the \rosat\ PSPC spectrum of the northern shell, 
but the eastern shell is well fitted with 
a single temperature model with a relatively low temperature of $kT =
0.21^{+0.05}_{-0.04}$~keV. They derive a shock velocity of
$v_{\rm s} = 380$~km~s$^{-1}$, which is much lower than the EPIC result, and 
a remnant age of $1.9 \times 10^{4}$~yr assuming a distance of 4~kpc. 
This age is 2 -- 4 times higher than our estimate. 
If we assumed a larger distance of 4~kpc, the calculation for the full 
equilibration case would yield $1.2 \times 10^{4}$~yr.

\section{Discussion}

\subsection{X-ray spectrum}

\rosat\ PSPC and Broad Band X-Ray Telescope (\bbxrt) spectra of \ctb\ have 
been analyzed \citet{1997ApJ...484..828R}. Assuming non-equilibrium ionization, 
they derive $kT$ = 1.7~keV and $n_{\rm e} t = 1.4\times 10^{10}$~cm$^{-3}$~s 
for the southern part of the remnant. A fit of the Lobe spectrum with a 
two-component Raymond \& Smith model yields 
$kT_{\rm low} = 0.16^{+0.03}_{-0.04}$~keV and 
$kT_{\rm high} = 0.56^{+0.20}_{-0.10}$~keV. Their estimate for the column 
density, \nh, for foreground absorption is 
$9.2^{+1.0}_{-1.5} \times 10^{21}$~cm$^{-2}$.
No spectral lines can be resolved in the \rosat\ PSPC spectrum, whereas the
improved spectral resolution of \xmmnewton\ EPIC allows lines to be identified,
and the continuum can also be better determined. Therefore,
with the \xmmnewton\ EPIC data we can show that the emission is arising from a
plasma out of ionization equilibrium and also determine the temperature more
accurately. In contrast to the \rosat\ results, we obtain a good fit of the 
EPIC spectra of the Lobe with a single NEI model with a temperature comparable 
to that of the high temperature component of \citet{1997ApJ...484..828R}. 
The column density \nh\ of the EPIC fit is lower than that from the 
\rosat\ data ($\sim5 \times 10^{21}$~cm$^{-2}$). 
Presumably, the lower spectral resolution data of \rosat\ PSPC allowed
successful fits to two equilibrium plasma models, but with a higher 
absorbing column density. 

\citet{1998A&A...330..175P} have fitted the \bepposax\ LECS spectrum of the 
whole remnant shell with a non-equilibrium ionization model and 
obtain $kT = 0.95^{+0.65}_{-0.27}$~keV,
$n_{\rm e} t = 3.8^{+3.8}_{-1.6} \times 10^{11}$~cm$^{-3}$~s, and 
\nh\ = $6.9^{+0.7}_{-1.2} \times 10^{21}$~cm$^{-2}$. The analyzed spectrum
does not include the emission from the pulsar and the Lobe, but encompasses
the rest of the remnant. The large errors of the fit parameters of the 
\bepposax\ spectrum are presumably caused by the lower spectral resolution 
compared to \xmmnewton\ EPIC as well as the large extraction region including 
emission from various parts of the SNR.

For the \xmmnewton\ EPIC data of \ctb\,, we use an NEI model for a collisional  
plasma with variable abundances for Mg and Si, while fixing the abundances of 
other elements to solar values.
In the shell and in the Lobe, the temperature is $kT = 0.50 - 0.70$~keV,
the ionization timescale $\tau \approx 1 - 3 \times 10^{11}$~s~cm$^{-3}$, and
the column density $\NH\ = 5 - 7 \times 10^{21}$~cm$^{-2}$, with the exception 
of region 5 in the northeastern shell, where the temperature is low ($\sim
0.3$~keV) and the ionization timescale is high 
($4 - 6 \times 10^{11}$~s~cm$^{-3}$).
In the region corresponding to the CO arm, the absorption seems to be higher 
with $\NH = (1.0\pm0.1) \times 10^{22}$~cm$^{-2}$, indicating that there is 
additional material along the line of sight to the SNR.
Abundances are consistent with solar values in the whole remnant, except for 
one region (the bright central region of the Lobe), where Si appears to be 
overabundant ($\sim1.6\pm0.2$ times solar). 
Since this marginal indication of overabundance is obtained only for
Si in one region within the Lobe, while all the other abundances are
consistent with solar, we conclude that there is no compelling 
evidence for ejecta emission. Given the large intensity variations
within the remnant, it is somewhat surprising that none of the
brighter regions shows evidence of enhanced abundances. It may be the
case in \ctb\ that the ejecta mass is relatively low and the current
amount of swept-up matter is relatively high such that the ejecta will
be difficult to detect. Although X-ray emission from ejecta is
generally observed in SNRs up to an age of a few $10^{3}$~yr, there are
remnants younger than \ctb\ which show no evidence for ejecta.
For example, Kes~79 is estimated to be $\sim 6000$~yr old and has
no obvious emission from ejecta \citep{2004ApJ...605..742S}.

\subsection{Interaction Between the SNR and the Giant Molecular Cloud}

As \citet{1985PASJ...37..345T} have shown, the CO emission stretches from the 
giant molecular cloud both north and south of \ctb\,. In the optical band, 
filamentary structures have been found that are located at the rim of faint 
\hii\ regions north and southeast of the SNR. At the position of the 
filaments, the \hii\ regions overlap with the remnant in projection.
This suggests that the optical emission is mainly caused by the 
interaction of the remnant with dense material harboring the \hii\ regions
\citep{1995AJ....110..747F}. The diffuse \hii\ regions are probably associated 
with the outskirts of the molecular cloud complex to the west.
From [\sii] line ratios, \citet{1995AJ....110..747F} derive pre-shock 
cloud densities of up to $\sim20$~cm$^{-3}$ for the optical filaments.
New \hi\ and CO data \citep{2002ApJ...576..169K} make it clear
that the remnant is located in a density gradient: very high density in the 
molecular cloud in the west and low density in the emission gap in the
east of the SNR. As \citet{2002ApJ...576..169K} point out, the \hi\ 
map indicates that the SNR is not expanding inside a stellar wind bubble, 
since there is no evidence for a lower density cavity in \hi\ with a 
pronounced rim.

\citet{1985A&A...145...70T} have performed two-dimensional numerical 
hydrodynamical calculations for the evolution of a supernova remnant in or 
near a molecular cloud. If a supernova with an initial energy of 
$E \approx 10^{51}$~ergs occurs in a low density interstellar medium 
(1~cm$^{-3}$), the remnant shell becomes radiative after 
$\sim 5 \times 10^{4} - 10^{5}$~yr 
when it has reached a radius of $R \approx 30$~pc. However, if the explosion 
occurs in a molecular cloud which has a higher density ($10^{3}$~cm$^{-3}$), 
the Sedov phase is shorter ($\sim 10^{3}$~yr) and $R$ smaller 
\citep[$\sim 1 - 6$~pc,][]
{1980ApJ...237..769S,1985A&A...145...70T,1999ApJ...511..798C}.
In the case of a supernova explosion inside a molecular cloud, a break-out 
occurs if the shock reaches the edge of the cloud. 
For \ctb\,, the position of the pulsar and the shell relative to the GMC 
complex as well as the semi-circular shape of the shell indicate that the 
supernova explosion was presumably located outside and close to a giant
molecular cloud. 
As \citet{1985A&A...145...70T} make clear in their calculations, a reflected 
shock forms from the cloud surface if the explosion takes place outside the 
cloud, but the shock transmitted into the cloud is weak and has a minor effect 
on the cloud surface. The reflected shock can cause an additional velocity 
component to the expansion of the SNR, directed away from the cloud. 
\citet{1992ApJ...388..127W} have modeled \ctb\ as a supernova explosion near 
(2~pc) a large molecular cloud, assuming an initial energy of 
$E = 3.6 \times 10^{50}$~ergs, an interstellar medium density of 
$n_{0} = 0.13$~cm$^{-3}$, and a cloud density of $n = 36$~cm$^{-3}$.
They reproduce a semi-circular shell of the observed size at the age of 
$1.3 \times 10^{4}$~yr. 

The deep EPIC mosaic image of \ctb\ confirms
that there is no emission in the western part of the remnant. In combination
with the morphology of the remnant in the radio, this indicates  
that the shock wave has been stopped completely in the west. 
However, no indication has been found for a molecular shock in the GMC
\citep[][and references therein]{1998AJ....116.1323K}. It is possible that
there is enough material in unshocked parts of the GMC which lie in front of
the interaction region such that any emission from this region could be
absorbed. 
Hard point sources discovered in the X-ray faint parts of the SNR 
(\S\ \ref{rgbima}) also suggest the existence of absorbing 
material in front of the SNR.

\subsection{Lobe as Shocked Cloud}

Our analysis of the \xmmnewton\ EPIC data shows that the spectral properties 
of the Lobe are very similar to those of the remnant shell. Moreover, the 
merged \xmmnewton\ image (Fig.\,\ref{color})
corroborates that the Lobe and the pulsar are not related morphologically, 
confirming that the Lobe is not a jet phenomenon.
\citet{1997ApJ...484..828R} have suggested that, due to the interaction 
between the SNR and the GMC in the west, a reflected shock has propagated 
into the SNR. Rayleigh-Taylor instabilities might have formed between the thin 
SNR plasma and the dense GMC, possibly producing a structure such as the Lobe.
Another possibility is that there are denser clouds in the outer parts of the 
GMC complex that have been shocked by the SNR blast wave. In projection, the 
Lobe and the bright knots in the northern part of the remnant seem to trace 
the outer boundary of the CO arm, and therefore might be emission from 
shocked interstellar clouds on the outskirts of the GMC complex.

\citet{1990ApJ...351..157T} point out that a density fluctuation in the  
pre-shock gas might have resulted in a local excess of the post-shock emission 
measure seen as the Lobe. They estimate that a cloud with a density 2 -- 5 
times higher than the average pre-shock density in the ambient medium could 
have produced the Lobe. 
However, they find no evidence of CO gas accelerated by the SNR shock wave.
In the far-infrared, \citet{1989MNRAS.238..649C} have found a region with 
bright emission northeast of the Lobe.
This region also coincides with the eastern edge of the CO arm extending from 
the western molecular cloud complex, which suggests
that this emission is caused by a shocked cloud.
\citet{1995AJ....110..747F} have found optical filaments at the southwestern 
edge of the Lobe, i.e.\ closer to the center of the remnant. All the other 
optical filaments detected in their observations are located along the 
northern and southern shell of the remnant and seem to be the result of the 
SNR interacting with material associated with the \hii\ regions in the north 
and the south. Usually, optical filaments 
are believed to be thin, shocked regions tangent to the line of sight. 
Since the diffuse optical emission in the north extends southward to
the position of the Lobe, they suggest that the central filamentary emission 
arises from shocked gas along the projected edge of an interstellar 
cloud. Assuming a shock velocity of 100~km~s$^{-1}$,
the observed [\sii] line ratio corresponds to a pre-shock cloud density of 
$n_{\rm cloud} = 5$~cm$^{-3}$ \citep{1995AJ....110..747F}. 

The interaction between a shock wave and a density enhancement like an 
interstellar cloud has been studied
theoretically by many authors. Two-dimensional hydrodynamical calculations
give information about the dynamical evolution of an interaction between the
SNR blast wave and a dense cloud \citep[e.g.][]{1990A&A...231..481B,
1994ApJ...433..757M,1994ApJ...420..213K}. The cloud fragments due to the
interaction. This is also observed in three-dimensional simulations by
\citet{1995ApJ...454..172X}, who in addition show that the morphology of the
cloud after the interaction depends strongly on the initial shape of the 
cloud. \citet{2003ApJ...583..245K} have performed a laboratory experiment using
the Nova high energy density laser at Lawrence Livermore National Laboratory
in order to study the interaction of a shock wave with a high-density
sphere located in a low-density medium. They confirm that the evolution of 
the sphere after the interaction depends on the density ratio of the 
high-density to low-density medium, $\chi$, and the Mach number of the shock 
wave, as shown by \citet{1994ApJ...420..213K}. The high-density sphere 
($\chi = 10$) is destroyed by the interaction with a shock wave with a Mach 
number of 10.

In order to better understand the possible origin of the Lobe emission in 
\ctb\, we estimate cloud densities from the results of \xmmnewton\ observations 
based on analytical estimates by \citet{1975ApJ...197..621S} for a planar 
shock-cloud interaction: There are two limiting cases, the `cold' cloud case 
and the `hot' cloud case, depending on the cooling time scale in the
post-shock cloud gas relative to the crossing time for the transmitted cloud 
shock. A critical density is estimated by equating the cooling time and the
shock crossing time:
\begin{equation}\label{critdens}
n_{\rm crit} = 5.8 \times 10^{-4} \beta^{\frac{5}{7}} v_{\rm s}^{\frac{10}{7}}
n_{0}^{\frac{5}{7}} a^{-\frac{2}{7}}, 
\end{equation}
where $\beta$ is the ratio of the pressure behind the transmitted cloud shock 
and the pressure in the shocked ambient ISM, $v_{\rm s}$ the shock velocity
(in km~s$^{-1}$), $n_{0}$ the ambient pre-shock density (in cm$^{-3}$), 
and $a$ the size of the cloud (in pc). For a density contrast of
$\chi = n_{\rm cloud}/n_{0}$ = 10, 100, and $\infty$, $\beta$ is 2.6, 4.4, and
6.0, respectively \citep{1975ApJ...197..621S}.
If the cloud is denser than $n_{\rm crit}$, the shocked cloud gas cools
rapidly (`cold' cloud) and, after having reached a temperature below 
$\sim10^{4}$~K, it produces line emission of lower ionization stages of the 
constituent elements. 
In the `hot' cloud case, the cloud density is lower than $n_{\rm crit}$
and the temperature is still higher than $\sim10^{6}$~K when the 
transmitted shock reaches the end of the cloud. The cloud emits X-rays and 
optical lines of higher ionization stages. However, there will be no region 
that has cooled down to $\sim10^{4}$~K, therefore no H$\alpha$ emission 
will be detected.

\begin{figure}
\centering
\includegraphics[width=\columnwidth]{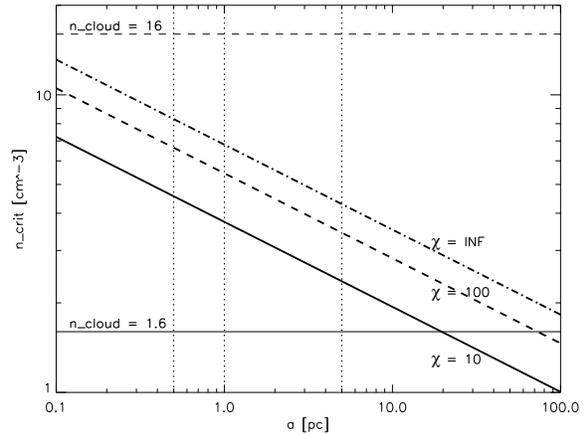}
\caption{\label{ncrit} 
{The critical density $n_{\rm crit}$ \citep{1975ApJ...197..621S} 
as a function of cloud size $a$ for
$v_{\rm s}$ = 720~km~s$^{-1}$ and $n_{0} = 0.16$~cm$^{-3}$ (thick lines). 
Solid line is used for $\chi$ = 10, dashed line for $\chi$ = 100, and 
dash-dotted line for $\chi = \infty$. For $\chi$ = 10, 100, the corresponding 
cloud density is shown with thin lines for $n_{0} = 0.16$~cm$^{-3}$. The dotted
lines indicate $a$ = 0.5, 1 and 5~pc.}
}
\end{figure}

To estimate the properties of a hypothetical uniform cloud in the Lobe region, 
we use the values obtained in \S\ \ref{sedovcalc} from the Sedov analysis 
of the EPIC data, i.e.\ $v_{\rm s} = 720$~km~s$^{-1}$ under the assumption of 
full equilibration 
between electrons and ions behind the shock, and $n_{0} = 0.16$~cm$^{-3}$. 
The critical density $n_{\rm crit}$ as a function of the cloud size $a$ 
(eq.\,(\ref{critdens})) is plotted in Figure \ref{ncrit} for three different
values for the density contrast $\chi = 10, 100, \infty$. The region below the
line corresponds to the `hot' cloud case, whereas the region above is the
`cold' cloud case. Assuming an ambient density of 
$n_{0} = 0.16$~cm$^{-3}$, 
the initial cloud density $n_{\rm cloud}$ = 1.6, 16 for $\chi = 10, 100$, 
respectively. $n_{\rm cloud}$ is also indicated in the diagram with thin
horizontal lines. 
As can be seen in Figure \ref{ncrit}, for $\chi = 100$ we would have
the `cold' cloud case for all clouds with $a > 0.1$~pc. However, the observed 
soft X-ray emission from the Lobe indicates a `hot' cloud case. 
If we assume a cloud size $a$ = 1~pc and a density contrast $\chi$ = 10, 
equation (\ref{critdens}) yields a critical density of 
$n_{\rm crit} = 3.7$~cm$^{-3}$, while a cloud size of $a$ = 0.5~pc with
$\chi$ = 10 results in $n_{\rm crit} = 4.6$~cm$^{-3}$.
For a likely cloud size $a \approx 1$~pc, $\chi = 10$ would result in a 
`hot' cloud. 
Therefore, the Lobe emission is indicative of an interaction between the shock 
wave of \ctb\ ($v_{\rm s} = 720$~km~s$^{-1}$, $n_{0} = 0.16$~cm$^{-3}$) and an 
interstellar cloud with a density $n_{\rm cloud} \la 5$~cm$^{-3}$.

\section{Summary}

We have studied the Galactic SNR \ctb\ using EPIC data of five \xmmnewton\ 
pointings. The deep EPIC image, created from all of these observations,
shows no emission in the western part of the shell.  This confirms
that the remnant has its semi-circular morphology because the progress
of the shock wave has been stopped by the GMC.  We find no
morphological evidence for a connection between the Lobe and the pulsar. 
The mosaic RGB image reveals some regions and clumps within the remnant which 
appear to have harder or softer spectra than the average spectrum. However,
a detailed spectral analysis indicates that only two regions have significantly
different fitted values for the \nh\,, temperature, or ionization timescale.
The Lobe region appears to have rather small spectral variations.

Using the RGB composite image as a guide, we extract spectra for 
different parts of the remnant in circular
regions as well as polygon regions. Best fit results are obtained with
a single-temperature non-equilibrium ionization model for a collisional plasma.
Abundances are determined for Mg and Si, while the abundances of the other
elements are fixed to solar values. 
As the spectra show no indication of nonthermal emission, we derive upper 
limits in the energy range of 0.3 -- 10.0~keV. In the eastern part of the
shell where emission is believed to arise from the forward shock the upper
limit of nonthermal emission is less than 2\% of the thermal emission. 

Analysis of the spectra of four regions in the Lobe indicates
that the spectral variations seen in the RGB composite image 
are caused by small differences in temperature, ionization timescale of the 
plasma, as well as \NH\,. None of these differences are significant
at the 90\% confidence level.  
There is no significant evidence for nonthermal emission from the Lobe. 
The upper limits of the nonthermal emission are low; the flux upper 
limit for the entire Lobe in the energy range of 0.3 -- 10.0~keV is 
$6.9 \times 10^{-14}$~ergs~cm$^{-2}$~s$^{-1}$
with an average surface brightness of 
$1.1 \times 10^{-15}$~ergs~cm$^{-2}$~s$^{-1}$~arcmin$^{-2}$,
about five times lower than the flux from the X-ray jet of the Vela pulsar.
The detailed spectral analysis shows that the spectra of the Lobe and 
the remnant shell are remarkably similar, with $kT = 0.50 - 0.70$~keV,
$\tau = 1- 3 \times 10^{11}$~s~cm$^{-3}$, and
$\NH\ = 5 - 7 \times 10^{21}$~cm$^{-2}$.  
Mg and Si abundances turn out to be consistent with solar values in almost
all analyzed regions of the SNR except in the bright central part of the Lobe 
where the Si abundance is slightly higher than solar ($\sim1.6\pm0.2$ times 
solar). 
This marginal indication of overabundance, only seen in one region of
the SNR, is not indicative of ejecta emission.
The small color variations seen in the RGB composite image do not result in 
significant differences in the fitted spectral parameters.
We use the Mg line emission as an NEI line diagnostic. 
We fit the \mgxi\ triplet and \mgxii\ Ly$\alpha$ line features with Gaussians 
and derive a lower limit for $kT$ which is consistent 
with the results from the global NEI fits of the EPIC spectra. 

The detailed spectral analysis does confirm significant spectral
variations for two locations within the remnant.
The northeastern part of the shell, right next to a radio bright knot,
contains a bright spot with lower temperature ($kT \approx 0.3$~keV) 
and higher ionization timescale ($\tau = 4 - 6 \times 10^{11}$~s~cm$^{-3}$).
In this region, \citet{1995AJ....110..747F} have found two bright optical 
filaments. The high value for $\tau$ indicates a region with a possibly 
higher density; it seems that the shock wave of the SNR has encountered a 
particularly dense cloud in this part of the shell.
The region just north of the pulsar corresponding to the CO arm
has a significantly higher \NH\/ ($\sim 1.0 \times 10^{22}$~cm$^{-2}$)
than other regions in the remnant.  The temperature and ionization
timescale are mostly consistent with the rest of the remnant.  
We conclude that the underlying spectrum in this region is not significantly 
different from the spectrum of the rest of the remnant, but the column 
density of material in front of this part
of the remnant is higher. We note that this larger value of \NH\/ is
consistent with what \citet{2003astroph0310575W} find 
for their fits to the spectrum
of 1E\,2259+586 and hence strengthens the already-strong case for an
association between the remnant and the pulsar.
In addition, there is marginal evidence for a spectral variation with
radius in the eastern shell of the SNR. The temperature is higher in the outer
eastern shell, whereas the \nh\ and $\tau$ values are higher for the inner
shell. This might indicate that the outer part of the shell (seen in 
projection) expands into a medium with lower density, while in the inner
regions the shock is affected by density enhancements in the ISM. 

From the result of the spectral analysis, we estimate an SNR blast wave 
velocity of $v_{\rm s} = 720\pm60$~km~s$^{-1}$ 
assuming that the remnant is in the Sedov phase 
and that there is full equilibration
of the electron and ion temperatures right behind the shock front, and a
remnant age of $t = (8.8\pm0.9) \times 10^{3}~d_{3}$~yr,
at an assumed distance of $D = 3.0~d_{3}$~kpc. In the case of partial 
equilibration with $T_{\rm e}/T_{\rm s} = 0.4$ as the lower limit, 
the blast wave velocity 
would be $v_{\rm s} = 1140\pm90$~km~s$^{-1}$ and the remnant age 
$t = (5.6\pm0.6) \times 10^{3}~d_{3}$~yr.
We also calculate a pre-shock density of the nuclei
$n_{0} = (0.16\pm0.02)~d_{3}^{-0.5}$~cm$^{-3}$, 
initial energy $E_{0} = (7.4\pm2.9) \times 10^{50}~d_{3}^{2.5}$~ergs 
(full equilibration) or $E_{0} = (18.5\pm7.3) \times
10^{50}~d_{3}^{2.5}$~ergs ($T_{\rm e}/T_{\rm s} = 0.4$), 
and swept-up mass of $M = (97\pm23)~d_{3}^{2.5}~M_{\sun}$.

The thermal nature of the X-ray emission of the Lobe as well as the optical 
filaments found at the southwestern edge of the Lobe are clear indications 
that the Lobe is the result of the SNR shock wave encountering
an interstellar cloud ($n_{\rm cloud} \la 5$~cm$^{-3}$). 
We infer from derived shock velocities and densities that the shock wave 
traveled through the cloud on a timescale which was comparable to the cooling 
time of the shocked gas in the cloud. The cloud was heated to temperatures
of $\sim10^{6}$~K and has not yet cooled down substantially, resulting in
little optical emission.

\acknowledgments

We are grateful to V.\ Kaspi for providing the \xmmnewton\ data (Obs.\ ID
0038140101) of 1E\,2259+586. We would like to thank A.\ Read and M.\ Freyberg
for useful discussion about the \xmmnewton\ EPIC background analysis. 
This work was supported by the NASA \xmmnewton\ grant NAG5-9914, 
NASA contracts NAS8-39073 and NAS8-03060, and in part by NASA grants GO0-1127X 
and GO1-2060X. This research has made use of the SIMBAD database,
operated at CDS, Strasbourg, France.

\appendix

\section{Calculating Non-equilibrium Line Emissivities}\label{neilineapp}

{\tt neiline} calculates line emissivities (using the Raymond \& Smith
\citeyear{1977ApJS...35..419R} code in its \citeyear{1993AAS...182.4127B} 
update) for certain non-equilibrium conditions,
specifically an ionizing or recombining plasma with astrophysical
abundances.  We used the code to calculate the line ratio diagnostics
described herein, and feel it could be of general interest to the SNR
community.

The code is quite simple.  After reading in the input values, it
initializes the Raymond \& Smith (\citeyear{1977ApJS...35..419R}; 
\citeyear{1993AAS...182.4127B} update) plasma code,
evolves the plasma, and outputs the requested emission. The user can
set the abundances, the initial electron and ionization
temperatures (the initial ionization temperature is used to set the initial 
ionization balance), the pressure, and the maximum time, ionization timescale, 
or minimum temperature for the model. The plasma evolution can be isothermal,
isobaric, isochoric, or a special case where the plasma is isobaric
until the temperature drops to a specified value after which it is
isothermal. Except in the isothermal case, the electron temperature drops
by a user-defined factor (default 2\%), and the cooling, ionization
balance evolution, and line emissivities are calculated by the Raymond
\& Smith code at each step in the evolution.

The code is written in C and Fortran, and is available at:\\
{\it http://cxc.harvard.edu/cont-soft/software/NEIline.1.00.html}\ .

\clearpage


\clearpage

\clearpage






\begin{table*}
\begin{center}
\caption{\label{obstab} \xmmnewton\ data used for the analysis.}
\begin{tabular}{ccccclc}
\tableline\tableline
\multicolumn{1}{c}{Obs.\ ID\tablenotemark{a}} & \multicolumn{2}{c}{Pointing direction} & 
\multicolumn{1}{c}{Inst.\tablenotemark{b}} & 
\multicolumn{1}{c}{Mode\tablenotemark{c}} & 
\multicolumn{1}{c}{Start time (UT)} & \multicolumn{1}{c}{Effective} \\
\multicolumn{1}{c}{} & \multicolumn{1}{c}{RA} &
\multicolumn{1}{c}{Dec} & \multicolumn{1}{c}{} & \multicolumn{1}{c}{} &
\multicolumn{1}{c}{} & \multicolumn{1}{c}{Exposure} \\
\multicolumn{1}{c}{} & \multicolumn{2}{c}{(J2000.0)} & 
\multicolumn{1}{c}{} & \multicolumn{1}{c}{} &
\multicolumn{1}{c}{} & \multicolumn{1}{c}{[ksec]} \\
\noalign{\smallskip}\tableline\noalign{\smallskip}                       
00575401 & 23 01 25.2 & 58 41 45 & PN & E.\ Full  & 
2002-01-22 17:03:34 & 8.0 
\\
(south, S) & & & M1 & Full  &  & 11.0
\\
 & & & M2 & Full  &  & 11.0
\\
00575402 & 23 00 48.8 & 59 05 27 & PN & E.\ Full & 
2002-07-09 07:58:03 & 6.0
\\
(north, N) & & & M1 & Full  &  & 11.0
\\
 & & & M2 & Full  &  & 11.0
\\
00575403 & 23 02 23.1 & 58 53 30 & PN & E.\ Full  & 
2002-07-09 13:34:48 & 9.0
\\
(east, E) & & & M1 & Full  &  & 11.0
\\
 & & & M2 & Full  &  & 12.0
\\
00381401 & 23 00 57.6 & 58 53 43 & M1 & Full  & 
2002-06-11 09:05:01 & 34.0
\\
(P1) & & & M2 & Full  &  & 34.0
\\
01553503 & 23 00 57.4 & 58 53 37 & M2 & Full  & 
2002-06-21 09:35:31 & 17.0
\\
(P2) & & & & & & 
\\
\noalign{\smallskip}\tableline
\end{tabular}
					
\tablenotetext{a}{Throughout the paper, 
the observations are called S, N, E, P1, and P2, as
indicated in the table.}
\tablenotetext{b}{Instruments: PN: EPIC PN, M1: EPIC MOS1, M2: EPIC MOS2.
All the analyzed data were obtained with the medium filter.}
\tablenotetext{c}{Full: Full Window Mode, E.\ Full: Extended Full Window Mode.}

\end{center}
\end{table*}

\begin{table*}
\centering
\caption[ ]{\label{mostab}
\xmmnewton\ spectral results of circular regions using the spectral model VNEI 
with solar abundances except for Mg and Si. Only results from the simultaneous 
fits of the MOS1 and MOS2 data are shown, whereas the paper discusses results
from PN data as well. Blank sky data are used as background data.
}
\begin{tabular}{lccccccc} 
\tableline\tableline 
Region & Pointing &\NH\/($10^{22}$) & $kT$ & Mg & Si & $nt (10^{11})$ &
$\chi^2$/d.o.f 
\\
& &[cm$^{-2}$] & [keV] &(solar) &(solar) & [s\,cm$^{-3}$] & 
\\
\tableline\noalign{\smallskip}
1 & P1 & 0.60$^{+0.02}_{-0.10}$ & 0.55$^{+0.04}_{-0.03}$ &
1.0$^{+0.2}_{-0.1}$ & 0.9$^{+0.2}_{-0.1}$ & 1.4$^{+1.1}_{-0.1}$ &
896.1/702 = 1.3
\\   
2 & P1 &  0.51$^{+0.05}_{-0.02}$ & 0.59$^{+0.02}_{-0.02}$ &
0.9$^{+0.1}_{-0.1}$ & 1.2$^{+0.2}_{-0.2}$ & 2.0$^{+0.1}_{-0.2}$ &
896.6/684 = 1.3
\\
3 & P1 & 0.49$^{+0.04}_{-0.03}$ & 0.55$^{+0.02}_{-0.01}$ &
0.9$^{+0.1}_{-0.1}$ & 1.6$^{+0.2}_{-0.2}$ & 2.0$^{+0.3}_{-0.2}$ &
904.3/721 = 1.3
\\    
4 & P1 & 0.49$^{+0.04}_{-0.05}$ & 0.53$^{+0.03}_{-0.01}$ &
0.9$^{+0.1}_{-0.1}$ & 1.2$^{+0.2}_{-0.2}$ & 2.8$^{+1.0}_{-0.5}$ &
896.5/676 = 1.3
\\
5 & E & 0.82$^{+0.07}_{-0.07}$ & 0.29$^{+0.03}_{-0.03}$ &
0.7$^{+0.2}_{-0.2}$ & 1.9$^{+0.9}_{-0.7}$ & 4.0$^{+4.1}_{-1.7}$ &
747.3/600 = 1.2
\\   
6 & E & 0.51$^{+0.07}_{-0.09}$ & 0.64$^{+0.09}_{-0.10}$ &
1.3$^{+0.4}_{-0.2}$ & 1.0$^{+0.3}_{-0.4}$ & 1.0$^{+0.6}_{-0.4}$ &
645.9/593 = 1.1
\\
7 & E & 0.68$^{+0.11}_{-0.06}$ & 0.63$^{+0.10}_{-0.11}$ &
1.1$^{+0.3}_{-0.2}$ & 1.0$^{+0.4}_{-0.4}$ & 0.8$^{+0.6}_{-0.3}$ &
678.0/552 = 1.2
\\
8 & E & 0.67$^{+0.03}_{-0.04}$ & 0.62$^{+0.04}_{-0.03}$ &
1.0$^{+0.3}_{-0.1}$ & 0.8$^{+0.2}_{-0.4}$ & 0.7$^{+0.2}_{-0.1}$ &
642.8/593 = 1.1
\\    
9 & E & 0.54$^{+0.49}_{-0.09}$ & 0.57$^{+0.11}_{-0.36}$ &
1.0$^{+0.6}_{-0.4}$ & 1.2$^{+0.8}_{-0.7}$ & 1.4$^{+1.4}_{-1.1}$ &
728.7/607 = 1.2
\\    
10 & P1 & $0.88^{+0.05}_{-0.02}$ & 
$0.52^{+0.01}_{-0.02}$ & $1.0^{+0.1}_{-0.2}$ & $1.0^{+0.3}_{-0.1}$ &
$1.7^{+0.5}_{-0.3}$ & 1126.7/918 = 1.2
\\
11 & N & 0.79$^{+0.22}_{-0.04}$ & 
0.53$^{+0.05}_{-0.24}$ & 0.8$^{+0.1}_{-0.2}$ & 0.6$^{+0.2}_{-0.2}$ & 
0.9$^{+0.5}_{-0.3}$ & 637.7/618 = 1.0
\\
12 & N & 0.70$^{+0.26}_{-0.11}$ & 0.48$^{+0.13}_{-0.23}$ &
1.0$^{+0.1}_{-0.4}$ & 1.0$^{+0.3}_{-0.3}$ & 1.1$^{+2.8}_{-0.5}$ &
653.4/586 = 1.1 
\\
13 & N & 0.67$^{+0.33}_{-0.05}$ & 0.52$^{+0.05}_{-0.26}$ &
1.0$^{+0.1}_{-0.1}$ & 0.9$^{+0.3}_{-0.2}$ & 1.4$^{+0.3}_{-0.4}$ &
733.9/617 = 1.2 
\\    
14 & N & 0.93$^{+0.14}_{-0.07}$ & 
0.42$^{+0.05}_{-0.13}$ & 0.8$^{+0.3}_{-0.4}$ & 0.8$^{+1.7}_{-0.2}$ & 
1.8$^{+N/A}_{-0.5}$ &
734.4/641 = 1.1 
\\    
15 & S & $0.54^{+0.01}_{-0.01}$ & $0.60^{+0.02}_{-0.01}$ 
& $1.0^{+0.1}_{-0.1}$ & $1.0^{+0.2}_{-0.3}$ &
$0.8^{+0.1}_{-0.2}$ & 1549.8/1325 = 1.2
\\
16 & S & $0.57^{+0.01}_{-0.01}$ & $0.60^{+0.03}_{-0.01}$ 
& $1.1^{+0.1}_{-0.1}$ & $0.9^{+0.3}_{-0.2}$ &
$1.0^{+0.1}_{-0.2}$ & 1646.6/1475 = 1.1
\\
\noalign{\smallskip}\tableline
\end{tabular}
\end{table*}

\begin{table*}
\centering
\caption[ ]{\label{evigwtab}
\xmmnewton\ spectral results of 
arc-shaped region in the eastern part of the shell, the CO arm, and the low 
surface brightness region in the southern interior, using {\tt evigweight}. 
The fit results are from the analysis of 
MOS1/2 data using the VNEI model; abundances are fixed to solar values 
except for Mg and Si. `Closed' data are used as background data.}
\begin{tabular}{lccccccc} 
\tableline\tableline 
Region & Pointing &\NH\/($10^{22}$) & $kT$ & Mg & Si & $nt (10^{11})$ &
$\chi^2$/d.o.f 
\\
& &[cm$^{-2}$] & [keV] &(solar) &(solar) & [s\,cm$^{-3}$] & 
\\
\tableline\noalign{\smallskip}
Outer E.\ Shell & E & 0.59$^{+0.02}_{-0.01}$ & 0.68$^{+0.02}_{-0.02}$ &
1.1$^{+0.2}_{-0.1}$ & 0.9$^{+0.2}_{-0.2}$ & 0.7$^{+0.1}_{-0.2}$ &
566.7/405 = 1.4
\\    
Inner E.\ Shell & E & 0.67$^{+0.04}_{-0.06}$ & 0.60$^{+0.05}_{-0.04}$ &
1.1$^{+0.1}_{-0.2}$ & 0.9$^{+0.2}_{-0.2}$ & 1.1$^{+0.3}_{-0.3}$ &
347.8/267 = 1.3
\\    
\noalign{\smallskip}\tableline\noalign{\smallskip}
CO Arm & P1 & $0.97^{+0.02}_{-0.01}$ & 
$0.54^{+0.01}_{-0.01}$ & $0.9^{+0.1}_{-0.2}$ & $0.9^{+0.2}_{-0.1}$ &
$1.5^{+0.4}_{-0.2}$ & 918.5/732 = 1.3 \\
\noalign{\smallskip}\tableline\noalign{\smallskip}
Dark Interior & P1 & $0.66^{+0.03}_{-0.07}$ & 
$0.53^{+0.05}_{-0.03}$ & $1.0^{+0.1}_{-0.2}$ & $0.9^{+0.1}_{-0.2}$ &
$1.5^{+0.4}_{-0.3}$ & 1452.5/1074 = 1.4 \\
\noalign{\smallskip}\tableline
\end{tabular}
\end{table*}

\begin{table*}
\centering
\caption[ ]{\label{mglineratio}
Results of the Mg line analysis from \xmmnewton\ EPIC MOS data 
(pointings P1 and P2) of the Lobe. 
The spectral model is `APEC No Line' plus Gaussians.}
\begin{tabular}{lcccc} 
\tableline\tableline
Region & \multicolumn{2}{c}{\mgxi\ triplet} & \mgxii\ Ly$\alpha$ & \mgxi\ triplet/\mgxii\ Ly$\alpha$ \\
& $E$ [keV] & Norm\tablenotemark{a} & Norm\tablenotemark{a} & Line Ratio \\
\tableline\noalign{\smallskip}
\multicolumn{5}{c}{P1, MOS1/2} \\
\noalign{\smallskip}\tableline\noalign{\smallskip}
1 (North) & $1.337_{-0.002}^{+0.009}$ & $1.1_{-0.5}^{+0.1} \times 10^{-4}$ 
& $2.0_{-0.5}^{+0.4} \times 10^{-5}$ & 
$5.5_{-3.0}^{+2.5}$ 
\\
2 (East) & $1.338_{-0.003}^{+0.008}$ & $9.7_{-2.0}^{+1.3} \times 10^{-5}$ 
& $0.1_{-0.1}^{+0.7} \times 10^{-5}$ &
$97_{-87}^{+\infty}$ 
\\
3 (Center) & $1.337_{-0.002}^{+0.004}$ & $1.2_{-0.2}^{+0.1} \times 10^{-4}$
& $7.7_{-4.0}^{+5.3} \times 10^{-6}$ &
$16_{-8}^{+19}$ 
\\
4 (Southwest) & $1.340_{-0.010}^{+0.006}$ & $7.2_{-2.0}^{+1.8} 
\times 10^{-5}$ 
& $0.1_{-0.1}^{+1.4} \times 10^{-5}$ &
$72_{-69}^{+\infty}$ 
\\
\noalign{\smallskip}\tableline\noalign{\smallskip}
\multicolumn{5}{c}{P2, MOS2} \\
\noalign{\smallskip}\tableline\noalign{\smallskip}
1 (North) & $1.335_{-0.010}^{+0.011}$ & $9.4_{-6.1}^{+2.6} \times 10^{-5}$ 
& $1.8_{-1.7}^{+0.8} \times 10^{-5}$ & 
$5.2_{-3.9}^{+114.8}$
\\
2 (East) & $1.343_{-0.008}^{+0.013}$ & $1.2_{-0.3}^{+0.2} \times 10^{-4}$ 
& $0.8_{-0.8}^{+1.8} \times 10^{-5}$ &
$15_{-12}^{+\infty}$ 
\\
3 (Center) & $1.335_{-0.005}^{+0.006}$ & $1.3_{-0.2}^{+0.5} \times 10^{-4}$
& $1.1_{-1.0}^{+0.7} \times 10^{-5}$ &
$12_{-7}^{+168}$ 
\\
4 (Southwest) & $1.340_{-0.010}^{+0.011}$ & $8.2_{-4.0}^{+1.8} 
\times 10^{-5}$ 
& $0.3_{-0.3}^{+0.8} \times 10^{-5}$ &
$27_{-23}^{+\infty}$ 
\\
\noalign{\smallskip}\tableline
\end{tabular}

\tablenotetext{a}{Norm = total number of photons/cm$^{2}$/s/extraction area 
in the line.}

\end{table*}

\end{document}